\documentclass[aps,pra,superscriptaddress,twocolumn,floatfix]{revtex4-1}
\usepackage[utf8]{inputenc}
\usepackage[T1]{fontenc}
\usepackage{amsmath}
\usepackage{amssymb}
\usepackage{mathtools}
\usepackage{graphicx}
\usepackage{overpic}
\usepackage[usenames,dvipsnames]{xcolor}
\usepackage{bm}
\usepackage[colorlinks,urlcolor=blue,citecolor=blue,linkcolor=blue]{hyperref}
\usepackage{url}
\usepackage{subfigure}
\usepackage{slashed,bbm}
\usepackage{dsfont}
\usepackage{setspace}
\usepackage{soul}
\usepackage{braket}
\usepackage{sidecap}
\usepackage{hyperref}
\usepackage{multirow}
\usepackage{tensor}

\usepackage[normalem]{ulem}
\usepackage{sidecap,tikz}
\DeclareRobustCommand{\orcidicon}{\hspace{-1.0mm}
	\begin{tikzpicture}
	\draw[lime, fill=lime] (0.0,0.0) 
	circle [radius=0.15] 
	node[white] {{\fontfamily{qag}\selectfont \tiny \,ID}};
	\draw[white, fill=white] (-0.0525,0.095) 
	circle [radius=0.007];
	\end{tikzpicture}
	\hspace{-3.0mm}
}
\foreach \x in {A, ..., Z}{\expandafter\xdef\csname orcid\x\endcsname{\noexpand\href{https://orcid.org/\csname orcidauthor\x\endcsname}
		{\noexpand\orcidicon}}
}

\definecolor{pink1}{RGB}{219, 48, 122}

\begin{document}

\title{Distinct quasiparticle interference patterns for surface impurity scattering on 
various Weyl semimetals}

\author{Feng Xiong}
\email{feng.xiong@rwth-aachen.de}
\affiliation{Institute for Theory of Statistical Physics, RWTH Aachen University, and JARA Fundamentals of Future Information Technology, 52062 Aachen, Germany}
\affiliation{Yangtze Delta Industrial Innovation Center of Quantum Science and Technology, Suzhou 215000, China}
\author{Chaocheng He}
\affiliation{Department of Physics, Zhejiang Normal University, Jinhua 321004, China}
\author{Yong Liu}
\affiliation{Yangtze Delta Industrial Innovation Center of Quantum Science and Technology, Suzhou 215000, China}
\author{Annica M. Black-Schaffer}
\affiliation{Department of Physics and Astronomy, Uppsala University, Box 516, 75120 Uppsala, Sweden}
\author{Tanay Nag\orcidB{}}
\email{tanay.nag@physics.uu.se}
\affiliation{Department of Physics and Astronomy, Uppsala University, Box 516, 75120 Uppsala, Sweden}

\date{\today}

\begin{abstract}
We examine the response of the Fermi arc in the context of quasi-particle interference (QPI) with regard to a localized surface impurity on various three-dimensional Weyl semimetals (WSMs). Our study also reveals the variation of the local density of states (LDOS), obtained by Fourier transforming the QPI profile, on the two-dimensional surface. We use the $T$-matrix formalism to numerically (analytically and numerically) capture the details of the momentum space scattering in QPI (real space decay in LDOS), considering relevant tight-binding lattice and/or low-energy continuum models modeling a range of different WSMs.   
In particular, we consider multi-WSM (mWSM), hosting multiple Fermi arcs between two opposite chirality Weyl nodes (WNs), where we find a universal $1/r$-decay ($r$ measuring the radial distance from the impurity core) of the impurity-induced LDOS, irrespective of the topological charge. Interestingly, the inter-Fermi arc scattering is only present for triple WSMs, where we find an additional $1/r^3$-decay as compared to double and single WSMs. The untilted single (double) [triple] WSM shows a straight-line (leaf-like) [oval-shaped] QPI profile. The above QPI profiles are canted for hybrid WSMs where type-I and type-II Weyl nodes coexist, however, hybrid single WSM demonstrates strong non-uniformity, unlike the hybrid double and triple WSMs. We also show that the chirality and the positions of the Weyl nodes imprint marked signatures in the QPI profile. This  allows us to distinguish between different WSMs, including the time-reversal-broken WSMs from the time-reversal-invariant WSM, even though both of the WSMs can host two pairs of Weyl nodes. 
Our study can thus shed light on experimentally obtainable complex QPI profiles and help differentiate different WSMs and their surface band structures.
\end{abstract}

\maketitle
\section{INTRODUCTION}

In condensed matter physics, the quantum Hall effect (QHE) introduces the concept of symmetry-protected phases that become relevant for topological quantum matter \cite{Klitzing80,Thouless82}. Digging deeper into topological matter, one can find that the bulk of the three-dimensional (3D) system can be gapless or gapped for Weyl semimetals (WSMs) \cite{Wan11,Burkov11,Aji12} and topological insulators \cite{xlQi11}, respectively. The former, namely WSMs, breaking either time-reversal symmetry (TRS), inversion symmetry (IS) or even both, hosts pairs of Weyl nodes (WNs) with opposite chirality \cite{yan2017topological} in momentum space. These WNs are considered to be the monopole and anti-monopole of the Berry curvature in reciprocal space.  A Fermi arc surface states exist between the surface projections of these bulk WNs. Additionally, a tilting term can tilt the conical bulk spectrum, resulting in a pocket-like Fermi surface instead of a point-like one around the WN, where the former and latter are known as type-I and type-II WNs, respectively \cite{YXu15,soluyanov2015type}. 
Depending on the different combinations of the WNs in the bulk, there exist two categories of WSM, namely pure and hybrid WSM. For pure WSMs, both WNs of the same pair are of same type, while for hybrid WSMs one WN  is type-I and its chiral partner belongs to type-II  \cite{fyli16,Shengyuan17,ALISULTANOV20183211,Mesot18,Nag22}.
Beyond the realm of single WSM with topological
charge $n=1$ and linear dispersion, it has been also shown theoretically that the anisotropic non-linear spectrum can lead to the topological charge $n$ greater than unity, referred to as multi-WSMs (mWSMs) \cite{Xu11,Fang12,Liu17,Ahn17,Roy17}. 
To be precise, $n=1$ $(n=2)$ $[n=3]$ is referred to as single (double) [triple] WSMs \cite{yang2014classification}.
The WSMs contribute significantly in the field of linear \cite{Lundgren14,Zyuzin17,Sharma16,Mukherjee17,Nandy19,Nag_2020a,sadhukhan2022effect,Feng22} and non-linear \cite{de2017quantized,Sadhukhan21,Sadhukhan21b,Das21,Nag22,Zeng21} transport phenomena due to their topological properties \cite{MZHasan17,potter2014quantum}, and chiral anomaly \cite{Huang15,Watzman18,li2016chiral,hirschberger2016chiral,Watzman18}. The WSMs have been experimentally realized in several inversion asymmetric compounds such as TaAs, MoTe$_2$, and WTe$_2$  (Co$_3$Sn$_2$S$_2$ , Heusler alloy-family, and rare
earth carbides-family)  \cite{lv2015observation,xu2015discovery,sadhukhan_npj,yang_17}. There also exist theoretical predictions about WSMs showing higher topological charges as well \cite{bernevig12,Liu17}. 

Having discussed the unique properties of WSMs, we note that there exist various theoretical techniques to probe their boundary effects other than exact diagonalization of tight-binding Hamiltonians such as, iterative methods to compute boundary Green’s functions \cite{morita1971lattice,Sancho_1984,Peng17}, and solving the Schrödinger equation \cite{Ningning08,Jian10,Delplace11,Falko13,Duncan18}. 
Moreover, impurities and defects in realistic materials 
can reveal the underlying topological nature of the boundary modes, which are considered to be experimentally relevant probes as well, in addition to the above theoretical techniques \cite{Zhou09,ran2009one,Hosur10,Liu09,Wei-Cheng09,Guo10,Black-Schaffer12,Fransson14}. For example, the impurity mediates scattering between different momentum modes leading to quasiparticle interference (QPI)
patterns. Interestingly, 
the interference between the incoming and outgoing momenta $k_{\rm in}$ and $k_{\rm out}$ associated with the impurity scattering can yield an amplitude modulation in the local density of states (LDOS) with spatial periodicity $2\pi/|k_{\rm in}-k_{\rm out}|$ \cite{mcelroy2003relating}. Such behavior, referred to as Friedel oscillations,
from which the Fermi momentum of the system under consideration can be estimated \cite{friedel1952xiv}. In the case of topological insulators, angle-resolved photoemission spectroscopy and scanning tunneling microscopy/spectroscopy (STM/STS) have been used to probe the topological nature of the surface states, following the QPI framework \cite{zhang2009experimental,roushan2009topological,Alpichshev10,zhu2018quasiparticle, TZhang09,Alpichshev10,beidenkopf2011spatial}.
Note that also graphene has paved to way to investigate the QPI from the theoretical side, as well as experimental perspective, following Fourier-transform (FT) STM techniques \cite{Cheianov06,Peres06,Skrypnyk06,Bena08,Dutreix16,Kaladzhyan21,Mallet07,MALLET2016294}. For example, it has been found that the LDOS decays algebraically for mono-layer graphene, exhibiting linear band touching, unlike other two-dimensional systems
\cite{Bena05a,peres2007electron,Bena08,Bena09}.


The theoretical computation of the QPI pattern for surfaces is possible within the $T$-matrix formalism \cite{Balatsky2006,Bena2016}, which is extremely useful in capturing the information about the boundary modes. With certain spatial profile of the impurity, analytical treatments with various types of Green's function are possible that yield a deeper physical insight than numerical techniques \cite{Kaladzhyan19,Pinon20b}. For instance, for infinite systems in one-, two- and three-dimensions, the impurity potentials are chosen to be point-like, line-like, and plane-like, respectively, such that the infinite systems split into two independent semi-infinite regions described by point-, line-, and surface- Green's functions. The decay profile of LDOS and oscillations, as obtained from this Green's function technique, depends on the nature of the impurity such as, magnetic or non-magnetic impurity and point-, or edge-like impurity \cite{JWang11,Biswas10,Biswas11,Alpichshev10}.

In recent times, the QPI profiles in WSMs have been studied to probe the boundary states, as well as bulk properties \cite{Mitchell16,Kourtis16,Lau17,Pinon20,Pinon20b}. Given the fact that the information about the surface Fermi arc states in WSMs can be extracted from the QPI profile in presence of an impurity plane, we here investigate the effect of a point impurity on the Fermi arc residing in a given surface of WSM. Our aim is to distinguish between single-, double-, triple- WSMs, and type-I, type-II, hybrid multi-WSMs, TRS invariant and broken WSMs by analyzing the interferences between incoming and outgoing wave packets, scattered by a single-point impurity. To be precise, we address the following questions that are experimentally pertinent as well: How do the higher topological charges influence the QPI profile and LDOS decay? What are the effects of the tilt? How is the interplay between opposite chirality Fermi arcs reflected there?


In this work, we consider a variety of WSMs where the nature and number of Fermi arcs are varied to investigate the QPI profiles both within the same Fermi arc and/or between two different Fermi arcs. We treat the problem with $T$-matrix formalism. 
We first analytically derive the decay signature of the LDOS with the radial distance $r$ from the core of the impurity, considering a low-energy surface Hamiltonian derived around the isolated WN that can also carry a higher topological charge (see Eqs.~(\ref{delr_n1}), (\ref{delr_n2}), (\ref{delr_n3})).  
We find that the single and double WSMs exhibit intra-Fermi arc scattering only where the LDOS decays as $1/r$.
By contrast, there is 
an additional $1/r^3$-decay in triple WSMs, where inter-Fermi arc scattering is observed. Still, the effects of higher topological charge are imprinted on the various powers of $1/r$-decay. Next, employing the $T$-matrix formalism on the tight-binding lattice models, we numerically compute both the QPI and LDOS profiles in the 2D momentum and real spaces, respectively,  of the WSM surface and as demonstrated in all the figures. Note that QPI and LDOS are related to each other by a 2D Fourier transformation.
We clearly observe line-, leaf-, and oval-shaped QPI profiles for untilted single, double, and triple WSMs, respectively (see Figs.~\ref{fig:sWSM}, \ref{fig:dWSM}, and \ref{fig:tWSM}) from which we can differentiate inter- and intra-Fermi arc scattering. 
The tilt further decorates the QPI patterns due to the presence of bulk modes in the Brillouin zone (BZ), allowing us to distinguish between the type-I and type-II WSMs. Hybrid WSMs, hosting both type-I and type-II WNs simultaneously, show a rotated QPI profile as compared to the type-I tilted WSMs (see Fig.~\ref{fig:hybrid}). Interestingly, the hybrid WSMs exhibit strong diagonal accumulation in the LDOS that is not observed for pure type-I or type-II WSMs. Turning our attention to comparing TRS broken and TRS invariant WSMs, now both hosting two pairs of WNs, we find that the inter-Fermi arc scattering, positions, and chiralities of the WNs can be understood by the distinct QPI profiles in terms of presence or absence of leaf-like patterns (see Fig.~\ref{fig:sWSMTIbroken}). On the other hand, the LDOS profiles for the above models look 
qualitatively similar, due to the fact that the inter and intra-Fermi arc scattering are simultaneously present in both models. 
The hybrid phase of the these models can be further distinguished by their QPI profile around the BZ boundary and angular LDOS structure (see Fig.~\ref{fig:sWSMTIbroken_hy}). Following this extensive analysis of various models, our results suggest that momentum space QPI and real space LDOS profiles are both useful in order to thoroughly differentiate WSMs.

The remainder of the work is organized as follows: In Sec.~\ref{QPIpatterns}, we discuss the $T$-matrix formalism for a single impurity on the top surface of a WSM. We here derive the numerical as well as analytical formalism. In Sec. \ref{analytical}, we report on the analytical findings for the impurity-induced LDOS.
In Sec. \ref{results}, we show the numerical results for the QPI and LDOS profiles for single, double, and triple WSMs to investigate the intra- and inter-Fermi arc scattering. We next compare the results for hybrid phases in these models. We then explore the QPI and LDOS profiles for two pairs of WNs and distinguish between TRS invariant and TRS broken single WSMs. Finally, we conclude with a discussion of possible future directions in Sec.~\ref{conclusion}.

\section{$T$-matrix method for a point impurity on the surface}
\label{QPIpatterns}

The Hamiltonian for WSMs can be generally written as $H= \sum_{\bm k} C^{\dagger}_{\bm k} h(\bm k) C_{\bm k} $ with $h(\bm k) =n_0 (\bm k) \sigma_0 +n_x(\bm k) \sigma_x + n_y(\bm k) \sigma_y+ n_z(\bm k) \sigma_z $, and $C({\bm k})= \begin{bmatrix} c_{{\bm k}\alpha }& c_{{\bm k}\beta} \end{bmatrix} ^T$ where $\alpha$, and $\beta$ denote the pseudo-spin degrees of freedom \cite{Armitage18,yan2017topological}. The specific forms of $n_i(\bm k)$ will be discussed in Sec.~\ref{analytical} and Sec. \ref{results} for low-energy and lattice models, respectively.  
Without loss of generality, we consider a simple case where the WNs only appear at ${\bm k}_c=(0,0, \pm k_{z_0})$ and the surface Fermi arc, connecting the projections of two WNs, in the $k_y$-$k_z$ ($k_x$-$k_z$) plane with open boundary condition along $x$ ($y$)-direction. For simplicity, the tilt term encoded in $n_0 (\bm k)$, only depends on $k_z$ that can characterize type-I and type-II WSM. 
Considering the open boundary condition along $y$-direction, the Hamiltonian can be written in terms of the good quantum momenta $\bm k_{\parallel}=(k_x,k_z)$ as it preserves the translational symmetry along $x$- and $z$-axis
\begin{align}
\hat{H}_0(\boldsymbol{k}_{\parallel})&=\sum^{L_y}_{y_n=1}
\sum_{l,\boldsymbol{k}_{\parallel}}\big[C^{\dagger}_{y_n}({\bm k}_{\parallel})h_{y_n,y_n}({\bm k}_{\parallel})C_{y_n}({\bm k}_{\parallel}) \nonumber \\
&+C^{\dagger}_{y_n}({\bm k}_{\parallel})h_{y_n,y_n+l}({\bm k}_{\parallel})C_{y_n+l}({\bm k}_{\parallel})+h.c.\big]\,.
\label{:eq:yslab}
\end{align} 
The first (second) term represents the on-site ($l$-th nearest order hopping) with $h_{i,j}({\bm k}_{\parallel})=n_0 (i,j,\bm k_{\parallel}) \sigma_0 +n_x (i,j,\bm k_{\parallel}) \sigma_x+n_y (i,j,\bm k_{\parallel})\sigma_y+n_z (i,j,\bm k_{\parallel}) \sigma_z$. $L_y$ is the length of lattice sites along the $y$-direction. 
Such a construction of the Hamiltonian is useful to analyze the problem of a single impurity $\hat{V}_{{\bm r_0}}$ localized on the top surface $y_n=1$ layer at ${\bm r}={\bm r_0}$. The strength of a scalar impurity is denoted by $V$, such that $\hat{V}_{{\bm r_0}}=V\delta({\bm r}- {\bm r_0})\sigma_0$. For this $(0,1,0)$ slab termination, a Fermi arc connects the surface projections of two bulk WNs in the pristine WSM.

The $T$-matrix formalism is a useful tool to study the effects of disorder in electronic systems using Green’s function method. It enables us to consider the contributions from all-order impurity scattering processes as long as the impurity is localized. On the other hand, one can use the Born approximation to examine the problem of extended impurities, under the condition of weak impurity scattering. By using the $T$-matrix technique, the full Green's function for $N$ number of localized impurities can be expressed as~\cite{Ziegler1996,Salkola1996,Mahan2000,Balatsky2006,Bena2016}
\begin{align}
\hat{G}(\boldsymbol{r}_i,\boldsymbol{r}_j,\omega)&= \hat{G}^0(\boldsymbol{r}_i,\boldsymbol{r}_j,\omega) \nonumber \\
&+\sum_{a,b=0}^{N-1}\hat{G}^0(\boldsymbol{r}_i,\boldsymbol{r}_a,\omega)\hat{T}(\boldsymbol{r}_a,\boldsymbol{r}_b,\omega)\hat{G}^0(\boldsymbol{r}_b,\boldsymbol{r}_j,\omega)\,, 
\label{:eq:fullG}
\end{align}
where $\hat{G}^0$ denotes the bare Green’s function for the clean system without impurities. The $T$-matrix obeys the Bethe-Salpeter equation
\begin{align}
\hat{T}(\boldsymbol{r}_a,\boldsymbol{r}_b,\omega)&=
\hat{V}_{\boldsymbol{r}_a} \delta({\boldsymbol{r}_a}- {\bm r_b}) \nonumber \\
&+ \hat{V}_{{\boldsymbol{r}_a}} \sum_{n=0}^{N-1} \hat{G}^0(\boldsymbol{r}_a,\boldsymbol{r}_n,\omega) \hat{T}(\boldsymbol{r}_n,\boldsymbol{r}_b,\omega)\,.
\label{:eq:Tmatrix}
\end{align}
In the present, we consider a single impurity i.e., $a=b=n=0$, located at $\boldsymbol{r}_0=(0,y_{n}=1,0)$ on the top surface. The onsite full Green's function at any position $\boldsymbol{r}=\boldsymbol{r}_i=\boldsymbol{r}_j$ 
is given by 
\begin{align}
 \hat{G}(\boldsymbol{r}_i,\boldsymbol{r}_i,\omega)= \hat{G}^0(\omega)+ \hat{G}^0(\boldsymbol{r}_i,\boldsymbol{r}_0,\omega)\hat{T}(\omega)\hat{G}^0(\boldsymbol{r}_0,\boldsymbol{r}_i,\omega)\,,
\label{:eq:fullGonsite} 
\end{align}
with $\hat{T}(\omega)=\big[\hat{I}-\hat{V}_{\boldsymbol{r}_0} \hat{G}^0(\omega)\big]^{-1}\hat{V}_{\boldsymbol{r}_0}$. Given the fact that the $\bm k_{\parallel}$ is a good quantum number for the bare Hamiltonian in Eq.~(\ref{:eq:yslab}), 
the bare Green's function without any impurity in momentum space can be obtained through $\hat{G}^{0}(\boldsymbol{k}_{\parallel},\omega)=\big[\omega+i \eta-\hat{H}_{0}(\boldsymbol{k}_{\parallel})\big]^{-1}$ with $\eta \to 0$. Further, the bare Green's function in real space can be calculated via FT of the Green's function in the momentum space,
\begin{align}
 \hat{G}^0(\boldsymbol{r}_i,\boldsymbol{r}_j,\omega)=\frac{1}{N_{\boldsymbol{k}_{\parallel}}}\sum_{\boldsymbol{k}_{\parallel}}\hat{G}^{0}(\boldsymbol{k}_{\parallel},\omega)e^{i\boldsymbol{k}_{\parallel}\cdot(\boldsymbol{r}_i-\boldsymbol{r}_j)}\,, 
\end{align}
where $\boldsymbol{r}_{i,j}=(x_{i,j},y_n=1,z_{i,j})$ denotes the spatial positions and $N_{\boldsymbol{k}_{\parallel}}$ is the number of $\boldsymbol{k}_{\parallel}$ points. As a result, for $\boldsymbol{r}_{i}=\boldsymbol{r}_{j}$, $\hat{G}^0(\omega)=\frac{1}{N_{\boldsymbol{k}_{\parallel}}}\sum_{\boldsymbol{k}_{\parallel}}\hat{G}^{0}(\boldsymbol{k}_{\parallel},\omega)$ represents the average value of $\hat{G}^{0}(\boldsymbol{k}_{\parallel},\omega)$ over whole momentum space. Since we are mainly interested in the variation of electronic densities on the surface, we can extract the LDOS through the onsite full Green's function at the top surface layer along the $y$ direction,
\begin{align}
\rho_{y_n=1}(\boldsymbol{r}_i,\omega)&=-\frac{1}{\pi}\text{Im}\bigg[\text{Tr}\big[G_{y_n=1}(\boldsymbol{r}_i,\boldsymbol{r}_i,\omega)\big] \bigg] \nonumber \\
&=\rho^{0}_{y_n=1}(\boldsymbol{r}_i,\omega)+\delta\rho_{y_n=1}(\boldsymbol{r}_i,\omega)\,.
\label{eq:ftqpi}
\end{align}
The first term $\rho^{0}_{y_n=1}$ is the LDOS in the clean system, while the latter term measures the fluctuation of LDOS induced by the impurity. By adopting a $\boldsymbol{r}$-mesh constrained on the top layer at $y_n=1$ with a grid size $N_{\boldsymbol{r}}=(N_x\times N_z)$, we can further FT the fluctuation part $\delta \rho$ back into the momentum space. The distribution of $\delta\rho_{y_n=1}(\boldsymbol{r}_i,\omega)$ in momentum space leads to the QPI for an isoenergy contour at $\omega$ given by
\begin{align}
 \delta\rho_{y_n=1}(\boldsymbol{q}_{\parallel},\omega)&=\rho_{y_n=1}(\boldsymbol{q}_{\parallel},\omega) - \rho^0_{y_n=1}(\boldsymbol{q}_{\parallel},\omega) \nonumber \\
 &=\text{Re}\bigg[\frac{1}{N_{\boldsymbol{r}}}\sum_{\boldsymbol{r}_i}\delta\rho_{y_n=1}(\boldsymbol{r}_i,\omega)e^{i\boldsymbol{q}_{\parallel}\cdot\boldsymbol{r}_i}\bigg]\,.
 \label{:eq:qpirho}
\end{align}
$\rho^{0}_{y_n=1}(\boldsymbol{q}_{\parallel},\omega=0)$ exhibits the
Fermi arc surface states as a signature of clean WSM. 

We here focus on pseudo-spin-integrated LDOS as a measure of the QPI profile. All the numerical results are obtained using a $400\times400$ grid in $\bm{q}_\parallel$-space and $\bm{r}_\parallel$ and with the infinitesimal imaginary part in the bare Green's function set to $\eta=10^{-4}$.
Note that the momentum modes are taken as $q_i=\pi m/N_i$ with $m\in (-N_i,N_i]$ where $N_i$ refers to the system size along $i$-direction.   The numerical result is independent of the number of   layers along $y$-direction due to the choice of the surface impurity.  
For the numerical calculations, we compute the changes $\delta\rho_{y_n=1}(q_x,q_z,\omega=0)$ [$\delta\rho_{y_n=1}(x,z,\omega=0)$] in the momentum [real] space to demonstrate the QPI [LDOS] profile, as reported in Sec.~\ref{results}, due to the localized point-impurity effect on the top surface.

We next cast the QPI profiles in momentum space for possible analytical solutions. The generalized Green’s function $\hat{G}(\boldsymbol{k}_1,\boldsymbol{k}_2,\omega)$ is related to the translationally invariant bare Green’s function $\hat{G}^0(\boldsymbol{k},\omega)$ as follows
\begin{align}
\hat{G}(\boldsymbol{k}_1,\boldsymbol{k}_2,\omega)&= \hat{G}^0(\boldsymbol{k}_1,\boldsymbol{k}_2,\omega) \delta(\boldsymbol{k}_1-\boldsymbol{k}_2)
\nonumber \\
&+ \hat{G}^0(\boldsymbol{k}_1,\omega)\hat{T}(\boldsymbol{k}_1,\boldsymbol{k}_2,\omega)\hat{G}^0(\boldsymbol{k}_2,\omega)\,.
\label{:eq:fullGk}
\end{align}
Note that the impurity breaks the translational invariance, resulting in the perturbed Green’s function dependent on two distinct momenta $\boldsymbol{k}_1$ and $\boldsymbol{k}_2$. The momentum $\bm{q}_\parallel$, discussed above, can be formally expressed as ${\bm q}_\parallel=\boldsymbol{k}_1-\boldsymbol{k}_2
\equiv \boldsymbol{k}_{\rm in} -\boldsymbol{k}_{\rm out}$. The $T$-matrix is thus found to be 
\begin{align}
\hat{T}(\boldsymbol{k}_1,\boldsymbol{k}_2,\omega)&=
\hat{V}({\boldsymbol{k}_1}, {\boldsymbol{k}_2}) \nonumber \\
&+\sum_{\boldsymbol{k}'} \hat{V}( {{\boldsymbol{k}_1}},{{\boldsymbol{k}'}}) \hat{G}^0(\boldsymbol{k}',\omega) \hat{T}(\boldsymbol{k}',\boldsymbol{k}_2,\omega)
\,.
\label{:eq:TKmatrix}
\end{align}

The exact solution for the $T$-matrix relies on the nature of the impurity considered. We here demonstrate the generic situations where the
impurity may reside in the bulk, over a plane, or on a line for the 3D system.
For the localized point impurity $\hat{V}_{{\bm r_0}}=V\delta({\bm r}- {\bm r_0})\sigma_0$ in the bulk of the system, $\hat{V}( {{\boldsymbol{k}_1}},{{\boldsymbol{k}'}})$ becomes independent of momentum. As a result, the $T$-matrix takes the form 
\begin{align}
\hat{T}(\omega)=[I - V \int \frac{d^3 \boldsymbol{k}}{\mathcal{V}} \hat{G}^0(\boldsymbol{k},\omega) ]^{-1} V 
\,,
\label{eq:TK1matrix}
\end{align}
where $\mathcal{V}$ denotes the volume of the BZ in 3D and the integral over $\boldsymbol{k}$ is performed over the entire BZ. The $T$-matrix does not contain any delta function as the translation symmetry is broken in all directions. For a line-like impurity $\hat{V}_{z}=V\delta(x- x_0) \delta(y- y_0)\sigma_0$ along the $z$-direction in 3D, the $T$-matrix contains a delta-function $\delta_{k1_z,k2_z}$ as the 
translation symmetry is preserved (broken) along the $z$- ($x$- and $y$-) direction. The $T$-matrix can hence be written as $\hat{T}(k_z,\omega)$ while $\int d^3 \boldsymbol{k}/\mathcal{V} \to \int dk_x dk_y/\mathcal{S}$, where $\mathcal{S}$ denotes the area of BZ along $k_x$-$k_y$ plane.
The case of plane-like impurity $\hat{V}_{yz}=V\delta(x- x_0)\sigma_0$ localized on the $yz$ surface at $x=x_0$, allows us to further reduce $\int d^3 \boldsymbol{k}/\mathcal{V} \to \int dk_x/\mathcal{L}$, where $\mathcal{L}$ denotes the width of BZ along $k_x$ only, with $\hat{T}(k_y,k_z,\omega)$, as the translational symmetry is preserved (broken) along $y$- and $z$- ($x$-) direction by the impurity along the $x$-direction. As a result, the $\hat{T}$ matrix contains the delta-function $\delta_{k1_y,k2_y} \delta_{k1_z,k2_z}$. Note that for 3D WSMs such plane-like impurity has already been used to analyze the Fermi arc surface states following this $T$-matrix formalism \cite{Pinon20,Pinon20b}. Our setup is different from this, as discussed below.  

In our case, with a surface point impurity at the top-most layer $y_n=1$, we compute the surface bare Green's function $\hat{G}^0(k_x,k_z,\omega)$ from the projected surface Hamiltonian $H(k_x,k_z)$ from the bare bulk Hamiltonian. As a result, $\hat{T}(\omega)$ is defined as 
\begin{align}
\hat{T}(\omega)=[I - V \int \frac{dk_x dk_z}{\mathcal{S}} \hat{G}^0(k_x,k_z,\omega) ]^{-1} V 
\,.
\label{eq:TK2matrix}
\end{align}
We can compute the onsite Green's function $\hat{G}^0(x,z,\omega)$ on the top surface using the FT of $\hat{G}^0(k_x,k_z,\omega)=[\omega + i\eta -\hat{H}(k_x,k_z)]^{-1}$. Here $\hat{H}(k_x,k_z)$ represents the surface Hamiltonian. The $T$-matrix does not depend on the momenta $\boldsymbol{k}_{1,2}$ enabling us to treat any $\boldsymbol{k}_{1}$- and $\boldsymbol{k}_{2}$-integral independently for $\hat{G}^0(\boldsymbol{k}_{1},\omega)$ and $\hat{G}^0(\boldsymbol{k}_{2},\omega)$ in Eq.~(\ref{:eq:fullGk}).
For the delta-function impurity located at $\boldsymbol{r}_0=(0,y_{n}=1,0)$ on the top surface, the modified LDOS at 
position $\boldsymbol{r}_i=(x_i,y_n=1,z_i)$
is then given by 
\begin{align}
 \delta \rho(\boldsymbol{r}_i,\omega)= -{\rm Im} [\hat{G}^0(\boldsymbol{r}_i-\boldsymbol{r}_0,\omega) \hat{T}(\omega) \hat{G}^0(\boldsymbol{r}_0-\boldsymbol{r}_i,\omega)].
\label{eq:LDOS_r} 
\end{align}
We are mainly interested in the quantity $\delta \rho $ induced by the impurity, demonstrated in the QPI (LDOS) profile studied in momentum (real) space. 
We note that Eq.~(\ref{eq:LDOS_r}) allows us to obtain an analytical expression of the LDOS considering the surface Green's function. Next, we analyze the radial variation of this LDOS.

\section{Analytical Results}
\label{analytical}

In order to analytically investigate the LDOS modulation due to an impurity on the top surface of a WSM, we start with a low-energy continuum model. 
The general low-energy form of a WSM around the WNs can be written as $H_{n}(\boldsymbol{k})= t_{0}k_z\sigma_{0}+ k_z\sigma_{z} +k^{n}_{-}\sigma_{+} +k^{n}_{+}\sigma_{-} $ where $\sigma_{\pm}=(\sigma_x \pm i \sigma_y)/2$, $k_{\pm}=k_x \pm i k_y$ and $n=1,2$, and $3$, represent single, double and triple WSMs, respectively. The type-I (type-II) phase is characterized by $t_0<1$ ($t_0>1$) with the tilt along the $k_z$-direction \cite{yang2014classification}.
The surface Hamiltonian $H^S_n \equiv \hat{H}(k_x,k_z)$, defined on the $xz$-surface, can be approximated by $H^{\rm S}_{n=1} = t_{0}k_z\sigma_{0}+ k_x \sigma_y + k_z \sigma_x$, $H^{\rm S}_{n=2}= t_{0}k_z\sigma_{0}+k_x^2 \sigma_z + k_z \sigma_x - k_x \sigma_y$ and $H^{\rm S}_{n=3}= t_{0}k_z\sigma_{0}+ k_x^2 \sigma_z + k_z \sigma_x + (k_x^3-k_x) \sigma_y$ for single, double, and triple WSMs, respectively \cite{Roy17,yang2014classification}. These surface Hamiltonians are obtained by projecting the bulk Hamiltonians $H_{n}(\boldsymbol{k})$ in the basis of $\sigma_y$, $\sigma_x$, $\sigma_y$ for $n=1$, $2$, and $3$ respectively, while the particular term $k_y^n$ is associated with the above Pauli matrices. Note that for open boundary condition along the $y$-direction, $k_y \to i\partial_y$. For simplicity we only consider the terms containing the maximum power of $(i\partial_y)$, given by $k^n_y $, to set up the basis for the surface projection. A more rigorous calculation would involve various powers of $(i\partial_y)$ that might yield a more accurate choice of the basis for such projection. As a result, the surface Hamiltonian can acquire a more complicated form than what we use and the above approach is thus only qualitative, with a more rigorous calculation for $H^{\rm S}_n$ left for future studies. 
We note that a point impurity on the top surface demands for such surface Hamiltonian, as far as our analytical method is concerned, the ability to extract the surface Green's function. Obtaining the surface Green's function
directly from the 3D bulk Hamiltonian for such surface impurity has so far been an open question.  Our approach is the first attempt, to the best of our knowledge, for solving the above problem from the analytical side.

We next validate the above approximation for surface Hamiltonians by corroborating the feature of the Fermi arcs obtained from a numerical lattice calculation. At the outset, we note that the analytical surface Hamiltonian is derived from a bulk Hamiltonian of an isolated WN. Therefore, it is not obvious to understand the Fermi arc quantitatively from this surface Hamiltonian. 
Technically we therefore need to consider two copies of such surface Hamiltonian in order to host a Fermi arc. Without loss of generality, we assume the Fermi arc between $-\pi/2<k_z<\pi/2$, with the WNs located at $(0,0,k_z=\pm \pi/2)$.
Nevertheless, we focus on the qualitative connection with just one WN as follows. 
For the untilted case, a single WSM, when solved by our numerical lattice calculations, hosts a straight Fermi arc, shown in Fig.~\ref{fig:sWSM}(a). This term is successfully predicted by the $k$-linear terms in $H^{\rm S}_{n=1}$. The bending of two Fermi arcs in double WSM, displayed in Fig.~\ref{fig:dWSM}(a), can be understood by the $k_x^2$ term in $H^{\rm S}_{n=2}$. In a similar spirit, two curved Fermi arcs along with a straight Fermi arc for a triple WSM, displayed in Fig.~\ref{fig:tWSM}(a), can be connected to the $k_x^2$ and $k_x$ terms in the $H^{\rm S}_{n=3}$, respectively. The number of Fermi arcs might thus be related to the maximum power of $k_x$. On the other hand, the effect of the tilt is also encoded in $H^{\rm S}_{n}$ where the tilt term with $k_z\sigma_0$ and Fermi arc related term with $k_z\sigma_x$ both exist. Below we continue to work with $H^{\rm S}_{n}$ to obtain $\delta \rho$, i.e.~the LDOS change from a single impurity in real space.

The surface bare Green's function for a single WSM is
\begin{widetext}
\begin{align}
\hat{G}^0_{n=1}(k_x,k_z,\omega)& =\frac{1}{\omega^2 -k^2} \begin{bmatrix}
\omega - t_0 k_z & k_z-i k_x \\
k_z+i k_x & \omega - t_0 k_z
\end{bmatrix}\,,
\label{G0_sWSM}
\end{align} 
while the surface Green's function for a double WSM is
\begin{align}
\hat{G}^0_{n=2}(k_x,k_z,\omega)&\equiv \frac{1}{\omega^2 -k^2 +{\mathcal O}(k^4)} \begin{bmatrix}
\omega - t_0 k_z + k_x^2 & k_z-i k_x \\
k_z+i k_x & \omega - t_0 k_z - k_x^2
\end{bmatrix}\,, 
\label{G0_dWSM}
\end{align} 
and the surface Green's function for a triple WSM is 
\begin{align}
\hat{G}^0_{n=3}(k_x,k_z,\omega)&\equiv \frac{1}{\omega^2 -k^2 +{\mathcal O}(k^4)} \begin{bmatrix}
\omega - t_0 k_z + k_x^2 & k_z-i (k_x^3-k_x) \\
k_z+i (k_x^3-k_x) & \omega - t_0 k_z - k_x^2
\end{bmatrix}\,. 
\label{G0_tWSM}
\end{align} 
\end{widetext}
In the following analytical treatment, we use polar coordinates such that 
$\int dk_x dk_z= \int k dk d\theta_k$ with $k=\sqrt{k_x^2+ k_z^2}$, $\theta={\rm arctan}(k_x/k_z)$.
The $T$-matrix following Eq.~(\ref{eq:TK2matrix}) becomes diagonal as the angular integration, defined on the 2D polar coordinate, for the off-diagonal terms vanish \textcolor{black}{$\int_0^{2\pi} \{\sin (l\theta_k),\cos (l\theta_k)\} \exp(i m \theta_k) d\theta_k =\{1,i\}l \delta_{lm}$}. The diagonal terms depend on $\omega^{-2} {}_2F_1(\Lambda,\omega)$ with $\Lambda$ being the upper cut-off for the $k$-integral and ${}_2F_1 $ denotes the hypergeometric function. For compactness, we refer to the diagonal terms in $\hat{T}(\omega)$ as $t(\omega)$. 
Using the Jacobi-Anger identity $\exp(i \boldsymbol{k}\cdot \boldsymbol{r})=\sum_l (i)^l J_l(kr) \exp(i l \theta_{kr})$ and following a FT to $\hat{G}(\boldsymbol{r},\omega)=\int d^2 \boldsymbol{k} \exp(i \boldsymbol{k}\cdot \boldsymbol{r} ) \hat{G}(\boldsymbol{k}, \omega)$ with $\boldsymbol{k}=(k_x,k_z)$, $\boldsymbol{r}=(r \sin \theta_r,r\cos \theta_r)$, and $\theta_{kr}=\theta_{k}-\theta_{r}$, we obtain a closed form expression of $\hat{G}(\boldsymbol{r},\omega)$ in terms of Bessel and hypergeometric functions. We are interested in the asymptotic limit $\omega r/v \gg 1$ (we consider the Fermi velocity $v$ to be unity for simplicity) to analyze the decay of the density with respect to $\boldsymbol{r}$ in real space. We find the following expressions of surface Green's function for single, double, and triple WSM:
\begin{widetext}
\begin{align}
\hat{G}^0_{n=1}(\boldsymbol{r},\omega)& \approx \cos (\omega r) \bigg (\frac{\omega}{r}\bigg)^{1/2} \begin{bmatrix}
1 - t_0 \exp(i \theta_r) & \exp(i \theta_r) \\
\exp(-i \theta_r) & 1 - t_0 \exp(i \theta_r)
\end{bmatrix}\,,
\label{G0r_sWSM}
\end{align}
\begin{align}
\hat{G}^0_{n=2}(\boldsymbol{r},\omega)& \approx \begin{bmatrix}
\cos (\omega r) (\frac{\omega}{r})^{1/2}(1 - t_0 \exp(i \theta_r)) + \frac{1}{\omega r^2} + f & \exp(i \theta_r) \cos (\omega r) (\frac{\omega}{r})^{1/2} \\
\exp(-i \theta_r) \cos (\omega r) (\frac{\omega}{r})^{1/2} & \cos (\omega r) (\frac{\omega}{r})^{1/2}(1 - t_0 \exp(i \theta_r)) + \frac{1}{\omega r^2} - f
\end{bmatrix}\,,
\label{G0r_dWSM}
\end{align} 
\begin{align}
\hat{G}^0_{n=3}(\boldsymbol{r},\omega)& \approx \begin{bmatrix}
\cos (\omega r) (\frac{\omega}{r})^{1/2}(1 - t_0 \exp(i \theta_r)) + \frac{1}{\omega r^2} + f & \exp(i \theta_r) \cos (\omega r) \frac{\omega^{1/2}}{r^{3/2}} - \frac{\omega^{5/2}}{r^{3/2}} \\
\exp(-i \theta_r) \cos (\omega r) \frac{\omega^{1/2}}{r^{3/2}} - \frac{\omega^{5/2}}{r^{3/2}} & \cos (\omega r) (\frac{\omega}{r})^{1/2}(1 - t_0 \exp(i \theta_r)) + \frac{1}{\omega r^2} - f
\end{bmatrix}\,,
\label{G0r_tWSM}
\end{align} 
\end{widetext}
with $f={\mathcal O}(\omega^{-2} r^{-4}) + {\mathcal O}(\frac{\omega^{3/2}}{r^{1/2}} \cos (\omega r))$.
During the above derivations of effective surface Green's functions, we only keep the Bessel functions to the $0$-th and $1$-st order. Note that $t_0(\omega/r)^{1/2}$ can become more important than ${\mathcal O}(\frac{\omega^{3/2}}{r^{1/2}}) \cos (\omega r)$ terms in the long-distance limit due to $\omega \to 0$ and $t_0 \ne 0$.

We can now compute the modification of the LDOS in real space analytically through \cite{bena2016friedel,Bena08,Bena09,Lounis11,weismann2009seeing}
\begin{widetext}
\begin{align}
\delta\rho_{y_n=1}(\boldsymbol{r},\omega) = -{\rm Im} [\hat{G}^0(\boldsymbol{r},\omega) \hat{T}(\omega) \hat{G}^0(-\boldsymbol{r},\omega)]
\approx - \sum_{\zeta,\zeta'} {\rm Im}[e^{i (\boldsymbol{k}_{\zeta'}- \boldsymbol{k}_\zeta)\cdot \boldsymbol{r}} \hat{G}^0_{\zeta}(\boldsymbol{r},\omega) \hat{T}(\omega) \hat{G}^0_{\zeta'}(-\boldsymbol{r},\omega)] = \sum_{\zeta,\zeta'} \delta\rho^{\zeta \zeta'}_{y_n=1}(\boldsymbol{r},\omega) \,,
\label{eq_delrho}
\end{align}
\end{widetext}
 giving rise to the QPI on the 2D surface, as a measure of scattering between the momentum modes $\zeta$ and $\zeta^{'}$ of WSMs.  Note that $\hat{G}^0_{\zeta}$ denotes the bare Green's function associated with  momentum mode $\boldsymbol{k}_\zeta$ while omitting the topological charge $n=1$, $n=2$, and $n=3$ for ease of the notation.
 At low-energies these modes are close to or on the surface Fermi arc depending on the tilt strength. Note that $\theta^{\zeta,\zeta^{'}}_r \to \theta^{\zeta,\zeta'}_r +\pi$ for $\boldsymbol{r} \to -\boldsymbol{r}$, where the ${\boldsymbol k}_\zeta \cdot {\boldsymbol r}$ term leads to the $\theta^{\zeta}_r$ angle. Note also that $G^0_{\zeta}= G^0_{\zeta'}$ as the surface Green's functions  in Eqs. (\ref{G0r_sWSM}-\ref{G0r_tWSM}) are defined on the real space ${\boldsymbol r}$ only. 
For simplicity of notation, we omit the subscript $y_n=1$ in $\delta\rho$ and we arrive at the following impurity-induced LDOS changes for single, double, and triple WSMs
\begin{widetext}
\begin{align}
 \delta\rho^{\zeta \zeta'}_{n=1}(\boldsymbol{r},\omega) \propto {\rm Re}\Bigg[ \frac{\omega }{r} t(\omega) \cos^2 (\omega r) e^{i (\boldsymbol{k}_{\zeta'}- \boldsymbol{k}_\zeta)\cdot \boldsymbol{r}} \bigg(1- e^{i (\theta_r^\zeta- \theta_r^{\zeta'})} -\chi \bigg) \Bigg]
 \label{delr_n1}
\end{align}
\begin{align}
 \delta\rho^{\zeta \zeta'}_{n=2}(\boldsymbol{r},\omega) \propto {\rm Re}\Bigg[ t(\omega) e^{i (\boldsymbol{k}_{\zeta'}- \boldsymbol{k}_\zeta)\cdot \boldsymbol{r}} \Bigg[ \frac{\omega}{r} \cos^2 (\omega r) \bigg(1- e^{i (\theta_r^\zeta- \theta_r^{\zeta'})}- \chi \bigg) + \frac{1}{\omega^2 r^4} + \frac{\cos (\omega r)}{\omega^{1/2} r^{5/2}} +{\mathcal O}(r^{-9/2}) + {\mathcal O}(t_0^2) \Bigg ] \Bigg ]
 \label{delr_n2}
\end{align}
\begin{align}
 \delta\rho^{\zeta \zeta'}_{n=3}(\boldsymbol{r},\omega) \propto {\rm Re}\Bigg[ t(\omega) e^{i (\boldsymbol{k}_{\zeta'}- \boldsymbol{k}_\zeta)\cdot \boldsymbol{r}} \Bigg[ \frac{\omega}{r} \cos^2 (\omega r) \bigg(1- \frac{e^{i (\theta_r^\zeta- \theta_r^{\zeta'})}}{r^2}- \chi \bigg) + \frac{1}{\omega^2 r^4} + \frac{\cos (\omega r)}{\omega^{1/2} r^{5/2}} +{\mathcal O}(r^{-9/2}) + {\mathcal O}(t_0^2) \Bigg ] \Bigg ],
 \label{delr_n3}
\end{align}
\end{widetext}
respectively,
with $\chi= t_0 (e^{i \theta_r^\zeta} - e^{i \theta_r^{\zeta'}})$ representing the effect of tilt. To continue we consider the following setup for the Fermi arc, generating the allowed momentum modes.
For untilted single WSM, $\boldsymbol{k}_{\zeta,\zeta'} \in (k_x=0,-\pi/2<k_z<\pi/2)$, while for small tilt, $\boldsymbol{k}_{\zeta,\zeta'} \in (0<k_x<k_x^c, -\pi/2<k_z<\pi/2)$, and for the overtilted case, bulk modes appear in addition to the Fermi arc even in the low-energy regime such that $\boldsymbol{k}_{\zeta,\zeta'} \in (-\pi<k_x<\pi, -\pi<k_z<\pi)$. For untilted double WSM, $\boldsymbol{k}_{\zeta,\zeta'} \in (-k_x^c<k_x \ne 0<k_x^c,-\pi/2<k_z<\pi/2)$, while for small tilt, $\boldsymbol{k}_{\zeta,\zeta'} \in (-k_x^{c1}<k_x\ne 0<k_x^{c2}, -\pi/2<k_z<\pi/2)$. For untilted and low-tilted triple WSM, the $\boldsymbol{k}_{\zeta,\zeta'}$ entails a similar set of momentum as observed for double WSM in addition to $\boldsymbol{k}_{\zeta,\zeta'} \in (k_x=0,-\pi/2<k_z<\pi/2)$ as found in single WSM. For the overtilted case in double and triple WSMs, $\boldsymbol{k}_{\zeta,\zeta'}$ behave similarly, however, there are additional momentum modes for triple WSM as compared to double WSM. The Fermi arcs are depicted in Fig.~\ref{fig:sWSM}(a), Fig.~\ref{fig:dWSM}(a), and Fig.~\ref{fig:tWSM}(a), respectively, for untilted single, double, and triple WSMs from which the above momentum intervals can be qualitatively motivated.  

We now analyze the functional dependence on distance $r$ from the impurity in the impurity-induced LDOS $\delta \rho$.  
For an untilted single WSM with $t_0=0$ in Eq.~\eqref{delr_n1}, it decays as $r^{-1}$ in addition to the sinusoidal variation as dictated by $e^{i (\boldsymbol{k}_{\zeta'}- \boldsymbol{k}_\zeta)\cdot \boldsymbol{r}} \cos^2 (\omega r)$. The momentum space profile of the Fermi arc determines the period of oscillations in the real space through the 
phase factor $e^{i (\boldsymbol{k}_{\zeta'}- \boldsymbol{k}_\zeta)\cdot \boldsymbol{r}}$ for $\omega \to 0$. 
Note that the angular profile is not isotropic rather LDOS builds up for $\theta^{\zeta,\zeta'}_r \approx \pm \pi/2$ as the Fermi arc lies along $k_z$ only. A close inspection suggests that $e^{i (\theta_r^\zeta- \theta_r^{\zeta'})} \to -1$ when $\theta^{\zeta}_r$ and $\theta^{\zeta'}_r$ are separated by $\pi$. The LDOS profiles along $x$-($z$-)direction are given by $\theta^{\zeta}_r \to \pm \pi/2~(0,\pi)$ where the spatial oscillations are governed by the above phase factors associated with $k^{\zeta}_x$ ($k^{\zeta}_z$).
For finite tilt $t_0\ne 0$, the angular profile will be visible for a finite region centered around $\theta^{\zeta,\zeta'}_r \approx \pm \pi/2$. As a result, $\chi \ne 0$, however, $e^{i (\theta_r^\zeta- \theta_r^{\zeta'})} \ne 1$, leading to a 
$\cos^2 (\omega r)/r$ behavior with $r=\sqrt{x^2 +y^2}$. We note that $\delta \rho$, according to Eq.~(\ref{delr_n1}), can vanish for the untilted case $\chi=0$ when $\theta^{\zeta}_r=\theta^{\zeta'}_r$. 
This results in a situation where LDOS profile becomes independent of $r$ in the leading order. Such an intriguing situation no longer arises even for infinitesimal tilt $\chi \ne 0$. We note that this $r$-independent behavior may be an artifact of the present method based on the effective surface Hamiltonian.

Based on Eq.~(\ref{delr_n2}), we discuss the following for a double WSM.
Without the tilt, the symmetric bending of the Fermi arc along $k_x$-direction would lead to a similar LDOS profile to that of a single WSM with tilt. The LDOS for double WSM has a similar $\cos^2 (\omega r)/r$ as the single WSM, which sets the longest distance decay. In addition, double WSMs host faster decaying terms, including a  $1/r^{4}$ term that is not accompanied by $\cos^2 (\omega r)$-variation. Overall this leads to a similar long-distance modulation for double WSM compared to single WSM, but with rapid short-distance decay. 
In presence of tilt, the asymmetric bending of Fermi arcs with $k_x$ leads to a more noticeable profile along $x$-direction. This is due to the fact that $e^{i (\theta_r^\zeta- \theta_r^{\zeta'})}$ could be $-1$ for certain $\theta$'s and $\chi \ne 0$. Unlike the single WSM, $\delta \rho$ continues to depend on $r$ even for untilted case with $\theta^{\zeta}_r=\theta^{\zeta'}_r$. This indicates that the LDOS always decays with $r$ for untilted double WSM. 

Finally, based on Eq.~(\ref{delr_n3}), we discuss triple WSMs. The untilted triple WSM shows an admixture of double and single WSM behavior as far as the decay profile is concerned. Surprisingly, the leading order $\cos^2(\omega r)/r$ decay is commonly observed irrespective of the charge of the WSMs, but we find that $\cos^2(\omega r)/r^3$ [$1/r^4$] decay is absent in double [single] WSM but present for triple WSM. 
For the bending nature of Fermi arcs, the LDOS shows an angular profile, however, mostly centered around $\theta^{\zeta,\zeta'}_r \approx \pm \pi/2$. The LDOS profile along $z$ at $x=0$ becomes further suppressed by $\chi$ factors as compared to the untilted case. 
Similar to the untilted double WSM, $\delta \rho$ continues to depend on $r$ even for untilted case. 
Notice that $\theta^{\zeta}_r=\theta^{\zeta'}_r$ results in an additional $1/r^3$ radial decay of LDOS in triple  WSM  irrespective of the tilt strength. 
In fact, we find a universal behavior for all the above WSMs in the overtilted case that is dominated by $\chi$ resulting in $\cos^{2}(\omega r)/r$ profile. All the above analytical findings for WSMs, in general can be useful to understand the numerical results based on a lattice model, that we discuss below in Sec.~\ref{results} for $\omega = 0$. Note however the above analysis was under the assumption that $\omega r \gg 1$ as we could not address the $\omega \rightarrow 0$ limit.

We next investigate the $r$-dependent algebraic terms in the LDOS and their possible origin as found in Eqs. (\ref{delr_n1}-\ref{delr_n3}). The exponential terms involving $(\boldsymbol{k}_{\zeta,\zeta'}\cdot \boldsymbol{r})$ and
$\theta^{\zeta,\zeta'}_r$ are modified when there exist inter- as well as intra-fermi arc scattering. In order to capture the inter-fermi arc scattering, $\theta_r^\zeta$ is always considered to be different from $\theta_r^{\zeta'}$ while for analyzing intra-fermi arc scattering, $\theta_r^\zeta$ and $\theta_r^{\zeta'}$ can be same as well as different. Note that $e^{i (\theta_r^\zeta- \theta_r^{\zeta'})} \cos^2(\omega r)/r$ term is present for all the WSMs irrespective of their topological charges; this could thus be tentatively considered as a signature of the intra-Fermi arc scattering that is present for single, double and triple WSMs. On the other hand, inter-Fermi arc scattering can be manifested in the term $e^{i (\theta_r^\zeta- \theta_r^{\zeta'})} \cos^2(\omega r)/r^3$ which is only present in triple WSM.
The linear $k_{x,z}$-terms in $H^S_n$ lead to $\cos^2(\omega r)/r$ behavior irrespective of the charge of the WSM. 
On the other hand, there exist $1/r^4$-decay for both double and triple WSM, unlike the single WSM which we believe can be considered to be a signature of non-linear dispersion. This might be related to the $k_x^2$-term present in the surface Hamiltonian for double and triple WSM. 
The $k_x^3$ term in triple WSM $H^S_{n=3}$ can potentially lead to the $\cos^2(\omega r)/r^3$ term which is absent in single and double WSMs. 
The above term-wise analysis is only qualitative and exact quantitative breakdown is mathematically extremely difficult due to the presence of various cross terms. We also emphasize that obtaining an accurate signature of inter-Fermi arc scattering through the radial variation of LDOS is beyond the scope of the present manuscript. We can only comment on some possible qualitative signatures. 
In our case of WSMs, inter-Fermi arc scattering might cause faster decay as compared to intra-Fermi arc scattering. 
On the other hand, the $1/r$ decay in bilayer graphene can be connected with that for the double WSMs as both the models are non-linear in momentum~\cite{Kaladzhyan21}.

The 2D FT of the LDOS $\delta\rho_{n}(\boldsymbol{r},\omega)$, expressed in Eqs.~(\ref{delr_n1}-\ref{delr_n3}), can yield 
the QPI profile in the momentum space. We can find $\exp(ilr)\cos^2(\omega r)/r \xrightarrow[\text{}]{\text{FT}} 2~ {\rm sign}[q+l] + {\rm sign}[q+l+2 \omega] + {\rm sign}[q+l-2 \omega]$, $\exp(ilr)/r^4 \xrightarrow[\text{}]{\text{FT}}(q+l)^3 {\rm sign}[q+l]$ and $\exp(ilr) \cos^2(\omega r)/r^3 \xrightarrow[\text{}]{\text{FT}} 2(q+l)^2 ~{\rm sign}[q+l] + (q+l+2 \omega)^2 {\rm sign}[q+l+2 \omega] + (q+l-2 \omega)^2 {\rm sign}[q+l-2 \omega])$ with $q=\sqrt{q_x^2+q_z^2}$ and $l$ corresponding to occupied momentum modes in the pristine system. The QPI profile becomes finite for $|q|< |l|$, provided $\omega \to 0$, where $q$ and $l$ can take positive and negative values.   
However, the different decay characteristics of LDOS with $r$ cause different shapes of QPI patterns as far as the algebraic $q$-dependence are concerned.

\textcolor{black}{To compliment the analytical results above, we also investigate another version of an effective low-energy model with the aim of see if a change could be advantageous for certain properties, while also investigating the robustness of our already derived results. In particular, the low-energy model considered above have only one isolated WN, and as such, the exponentially localized surface states at zero energy, reached by applying open boundary condition along $y$-direction,  is not evident when doing the substitution $k_y \to -i \partial_{y}$.  If we instead concider an alternative effective low-energy model with the $k_z\sigma_z$ term replaced with $(\alpha-k_x^2-k_y^2-k_z^2) \sigma_z$ keeping the rest unaltered, we recover the two WNs, located at $(0,0,k_z)=(0,0,\pm \sqrt{\alpha})$. Introducing a surface along $y$-direction we arrive at $H(k_x,\partial_y,k_x)=H_0(\partial_y) + H_1(k_x,k_z)$
with $H_0(\partial_y)=(\alpha-\partial^2_y)\sigma_z -i \partial_y \sigma_y$, $H_1(k_x,k_z)=-(k_x^2+k_z^2) \sigma_z+ k_x \sigma_x $. This allows us to obtain 
the exponentially localized surface states at zero energy $\psi \sim N(\alpha)\exp(-y/2 +i k_x x + i k_z z) (1,1)^T $ such that $H_0\psi=0$, where $N(\alpha)$ is a normalization factor.  This solution can be anticipated to form the surface Fermi arc in the $k_xk_z$-plane  while the underlying bulk Hamiltonian now encompasses the information of two distinct and well separated WNs. As such this is a more realistic model of both the bulk and the surface than our previously employed analytical model.
However, the Hamiltonian for the surface Fermi arc can still not be derived  from the projection $\langle \chi |H_1|\chi \rangle$ with $|\chi\rangle=(1,1)^T$. In fact, the derivation of the  surface Hamiltonian, mimicking the Fermi arc, is an open problem, while the boundary projected Hamiltonian are obtained for other cases \cite{Ya-Jie20,Ghosh21a}. This alternative effective low-energy bulk Hamiltonian hence does not have any real additional advantage over the previous one as far as the computation of the surface Hamiltonian and mimicking the Fermi arc, is concerned.
} 

\textcolor{black}{Still proceeding with our alternative low-energy model, we next consider the change it induces to the radial profile of the LDOS in Eqs.~(\ref{delr_n1}), (\ref{delr_n2}), and (\ref{delr_n3}). A detailed derivation gives that the $k_z$ terms, appearing on the off-diagonal parts of the surface bare Green's function $\hat{G}^0_{n}(k_x,k_z,\omega)$ in Eqs.~(\ref{G0_sWSM}), (\ref{G0_dWSM}), and (\ref{G0_tWSM}), are replaced by  $\alpha-k^2$ where $k^2=k_x^2+k_z^2$.The off-diagonal part of the real space surface Green's function $\hat{G}^0_{n}(\boldsymbol{r},\omega)$, obtained by Fourier transformation, consists of $r^{-1/2}$. This is due to the fact that the leading order contribution of $\int k(\alpha-k^2)J_0(kr)dk/(\omega^2-k^2)$ yields $ \alpha/(\omega r)^{1/2} +\omega^{3/2}/r^{1/2}$. This  causes $r^{-1}$ behavior of $\delta \rho$ when the $(\alpha-k^2)\sigma_z$  term is considered.
Note that this $r^{-1}$-dependence is already present in our previous analysis for impurity-induced LDOS as shown in Eqs.~(20), (21), and (22) using the $k_z\sigma_z$ term. 
Therefore, even after taking into account the $(\alpha-k^2)\sigma_z$  term, we do not expect any qualitative change in the impurity-induced LDOS, as far as the linear radial decay is concerned. There might exist other $\omega$-dependent terms different from the linear $\omega$ terms in $\delta \rho$, but since we are interested in the functional form of $\delta \rho$ in terms of $r$ mainly, we can still comment that the $1/r$ decay seems to be unambiguously predicted from low-energy models and appears both for a single WN as well as models where two WNs are present.}

\textcolor{black}{As a final comment, we note that surface Green's functions in Eqs.~(\ref{G0_sWSM}), (\ref{G0_dWSM}), and (\ref{G0_tWSM}) are not able to mimic the  Fermi arc accurately extending between two discrete surface projections of the bulk WNs. This is true both for the low-energy models having a single WN and the alternative model with two WNs. 
It would be possible in principle to capture the surface Fermi arc from the bulk Weyl nodes (WNs) by considering a plane of impurities leading to an open surface \cite{Pinon20b}. However, in such an approach, one would need to use the a $T$-matrix formalism at the beginning in order to get the surface Green's function, carrying the information of the surface Fermi arc. To be precise, one can introduce a plane of impurities in the $xz$-plane breaking the translation symmetry along the $y$-axis only. This would then lead to the surface Green's function hosting Fermi arc on the $k_xk_z$-plane. We do not choose to follow this approach as we need then adopt two successive $T$-matrix operations to look for the impurity scattering  within and/or among Fermi arcs in the presence of a single-point impurity. The first $T$-matrix is for obtaining the surface Green's function from the bulk WNs and the second $T$-matrix is for impurity-induced LDOS  from the surface Green's function. This would be too complicated to handle analytically within the scope of this work.
Having said that, we still expect the algebraic $1/r$-decay to be present also in calculations that could capture the Fermi arc, as the Bessel function $J_0(kr)$ is the most significant  one among all $J_n$'s. Apart from the this decay due to intra-Fermi arc scattering, the dominant contribution for intra-Fermi arc scattering leading to $1/r^3$-decay is also caused by  $J_0(kr)$.
To summarize, while the analytical information we derive in this section is approximative and miss some essential facets of the problem, it is still useful in providing some analysis of the full lattice model numerical results we present in the following Sec.~\ref{results}. At the same time we would like to note that analytical results are derived with small but finite $\omega \ne 0$ while numerical results are for $\omega = 0$. }

\section{Numerical Results} \label{results}
We next numerically examine the lattice results for various WSMs as described below. 
\textcolor{black}{These numerical results involve the full 3D lattice of the WSM and hence accurately capture both all WNs and the Fermi arc(s) connecting them. Also, these numerical results are performed independently of the analytical results in the previous section, however, whenever beneficial we connect these analytical results.}

\subsection{Single WSMs}
\label{res1}

\begin{figure}[ht!]
	\begin{center}
		\includegraphics[width=1.0\linewidth]{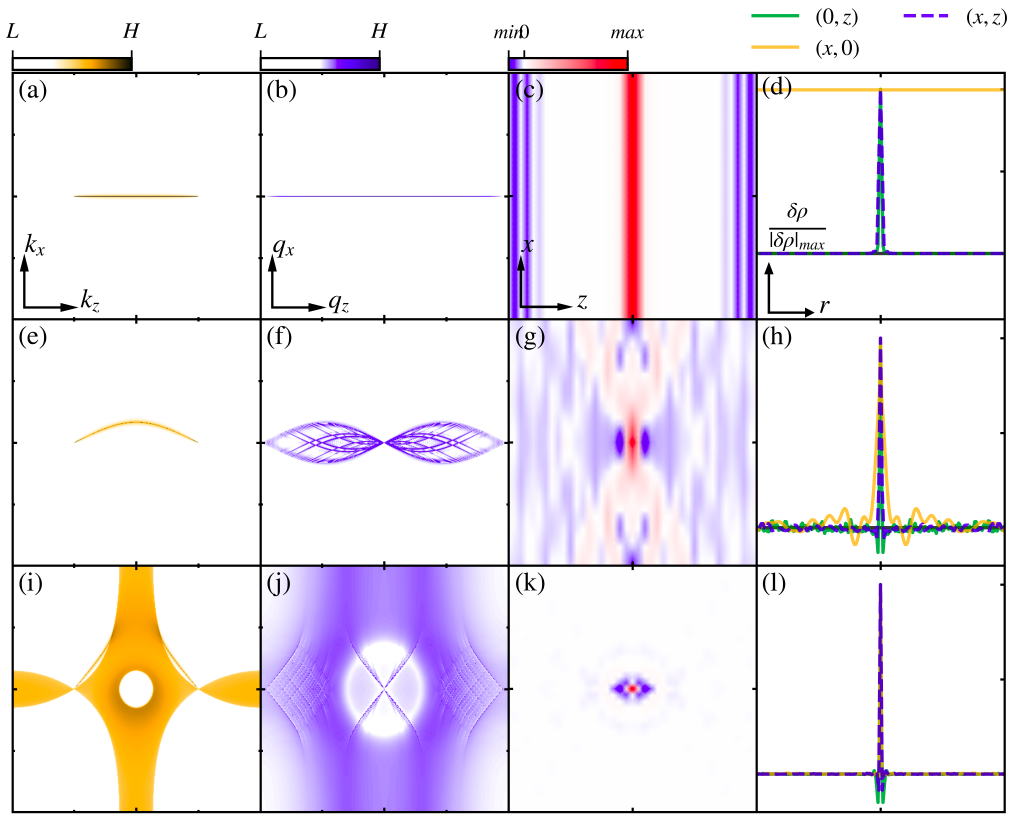}	
	\end{center}
	\vspace{-0.4cm}
	\caption{Single WSM: Momentum resolved surface spectral weight (yellow)	$\rho^0_{y_n=1}(\boldsymbol{k}_{\parallel},\omega)$ for clean WSM and impurity-mediated QPI profile (blue) $\delta\rho_{y_n=1}(\boldsymbol{q}_{\parallel},\omega)$, obtained from Eq.~(\ref{:eq:qpirho}), are demonstrated in (a), (e), (i), and (b), (f), (j), respectively. The LDOS change, or the Fourier transformed (FT) QPI, in Eq.~(\ref{eq:ftqpi}) on the top $xz$-surface that embeds the localized impurity at $(x,z)=(0,0)$ in (c), (g), and (k). The normalized radial decay profiles are depicted in (d), (h), and (l), qualitatively corroborating with the analytical findings in Eq.~(\ref{delr_n1}). We choose tilt strengths $t_0=0$, $0.5$, and $1.2$ for upper (a), (b), (c), (d) (untilted WNs), middle (e), (f), (g), (h) (type-I WNs), and lower (i), (j), (k), (l) (type-II WNs) panels, respectively. We choose $\omega= 0$ to look for scattering phenomena caused by the zero-energy surface states. The axes for the plots are shown in the lower left corner in the top row and are then used for all subsequent plots in the same column. We consider $-\pi< k_x, k_z,q_x,q_z \le \pi$  for the momentum  space resolved LDOS in the first and second  columns. In the third column, we show the microscopic variation of LDOS for  $-20\le x,z\le 20$.
 In the fourth column, we choose $r\in[-100,100]$ to study the macroscopic radial behavior. Note that $r=\sqrt{x^2+z^2}$, while $\theta_r=0$, $\pi/2$ and $\pi/4$ correspond to $(0,z)$, $(x,0)$, and $(x,z)$, respectively. We consider $[-200:200]\times [-200:200]$-grid for the top $xz$ layer in momentum as well as real space. 
 Note that the third column captures the LDOS within a short spatial range around the impurity site, while the fourth column apprehends the variation of LDOS for a longer range. The red (blue) color corresponds to the accumulation (depletion) of LDOS in the third column. We consider impurity strength $V=1$ in the unit of the hopping $t=1$ for single WSMs. We adopt the above representation in all the following figures.
 For single WSMs we find that  with increasing tilt, the Fermi arc bends and the QPI profile acquires a leaf-like structure. The inhomogeneous LDOS structure becomes prominent for intermediate tilt strength, while the LDOS is highly anisotropic (isotropic) for the untilted (overtilted) case.
	} \label{fig:sWSM}
\end{figure}

For single WSMs we adopt $(n_0,n_x,n_y,n_z)=(t_0 \cos k_z,\sin k_x, \sin k_y, m_z-\cos k_x-\cos k_y-\cos k_z)$ where the WNs of topological charge $n=1$ with chirality $\pm 1$ are located at ${\bm k}_c=(0,0,\pm \pi/2)$ for $m_z=2$~\cite{McCormick17,Roy17,Nag20}. For the untilted case with $t_0=0$ the surface Fermi arc with momenta ${\bm k}_{\rm FA}$ is a straight line with $k^x_{\rm FA}=0$ (see Fig.~\ref{fig:sWSM}(a)). The most favorable scattering momenta, arising from ${\bm k}_{\rm in}, {\bm k}_{\rm out} \in {\bm k}_{\rm FA}$, thus share an identical $k_x$ component as the Fermi arc does not disperse along $k_x$. The resulting QPI pattern does not have any dispersion along $q_x$ as ${\bm k}^x_{\rm in}={\bm k}^x_{\rm out}=0$ while it only stretches along $q_z$ due to the fact that ${\bm k}^z_{\rm in}\ne {\bm k}^z_{\rm out} $ (see Fig.~\ref{fig:sWSM}(b)). Notably, the intensity for $q_z=\pm \pi$ in the QPI profile is substantially weaker compared to the rest of the $q_z$'s. The resulting LDOS change thus is vanishingly small around the surface projection of the bulk WNs at $k_z=\pm \pi/2$ over the Fermi arc. 
As the tilt increases, the Fermi arc disperses along the $k_x$ direction, causing a narrow belt of momenta with $k_x\ne 0$ to contribute into the scattering process (see Fig.~\ref{fig:sWSM}(e)). For the tilted type-I case at $t_0=0.5$, the curved Fermi arc displays a leaf-like QPI profile where the width along $q_{x (z)}$-direction is determined by $q_{x,z}=|{\rm min}\{ {\bm k}^{x,z}_{\rm in} \} - {\rm max}\{ {\bm k}^{x,z}_{\rm out} \}|$ (see Fig.~\ref{fig:sWSM}(f)). The leaf-like structure is oriented along $z$-direction as the Fermi arc spans between $k_z=\pm \pi/2$, referred to as a ``figure 8" in a previous study~\cite{Kourtis16}, although the evolution of such leaf-like pattern was not studied under the gradual application of tilt. For the overtilted type-II case of $t_0=1.2$, we find that the Fermi arc surface states dissolve into the bulk modes with substantially extended electron and hole pockets (see Fig.~\ref{fig:sWSM}(i)). The "X"-shaped structure of QPI near the origin $(q_x,q_z)=(0,0)$ can be simply viewed as the intra-Fermi arc scattering (see Fig.~\ref{fig:sWSM}(j)). However, additional scattering from the bulk modes leads to a blurred background in the QPI profile, making intra-Fermi arc scattering less visible compared to untilted or type-I WSMs.

We next concentrate on the spatial variation of LDOS with impurities. The impurity can lead to prominent accumulation or weak depletion of LDOS as shown by red and blue color in Fig.~\ref{fig:sWSM}(c,g,k). Overall, the depletion is more spread out while accumulation is concentrated.  For untilted WSMs, the LDOS shows a $1/z$ decay along $z$-direction around $x=0$ while it behaves as constant along the $x$-direction around $z=0$ (see Fig.~\ref{fig:sWSM}(c)). 
The decay profile along $z$($x$)-direction is suppressed (not suppressed) substantiating the anisotropic nature of LDOS around the impurity core that is consistent with the anisotropic QPI profile (see Fig.~\ref{fig:sWSM}(d)). We note that the $r$-independent behavior of LDOS along $z$-direction at $x=0$ in untilted single WSM can be qualitatively understood from the analytical findings with $\theta^{\zeta}_r=\theta^{\zeta'}_r$ and $\chi=0$ when $\delta \rho $ becomes $r$-independent in the leading order (see Eq.~(\ref{delr_n1})). We attribute this behavior to the idealized straight Fermi arc with its completely flat energy dispersion at zero-energy. This generates an idealized  QPI profile localized at $q_x=0$ and varying only with $q_z$, which then results in this idealized real space behavior. In a real material, the Fermi arc no longer remains completely straight will be dispersive, which induces a decay in the LDOS. In the present case, depletion in LDOS can occur for $\omega \ne0$ such that its accumulation at $\omega =0$ is compensated.    

For the tilted case, the leaf-like QPI structure of finite width along $q_x$ yields a variation of the LDOS with $x$ over the $z=0$-line (see Fig.~\ref{fig:sWSM}(g,k)). Such angular variation is caused by the $\exp(i\theta^{\zeta}_r)$ term, however, the modulation decays with distance from the center (see Fig.~\ref{fig:sWSM}(h)). 
The short-(long-)period oscillations along $z$-($x$-)direction can be understood analytically from $\delta\rho$ (Eq.~(\ref{delr_n1})), where the phase factor $ e^{i (\boldsymbol{k}_{\zeta'}- \boldsymbol{k}_\zeta)\cdot \boldsymbol{r}}$ with the available momenta $\boldsymbol{k}_{\zeta,\zeta'}\in {\bm k}_{\rm in}, {\bm k}_{\rm out}$ bears the signature of spatial periodicity. 
For the overtilted case, the LDOS decays rapidly (uniformly) with $r$ (angle) as dominated by the tilt term (see Fig.~\ref{fig:sWSM}(l)). A qualitatively isotropic LDOS profile, as compared to the previous anisotropic behavior under small tilt, is observed for the overtilted case (see Fig.~\ref{fig:sWSM}(k)). This might be caused by the uniform variation of phase factors $\theta^{\zeta,\zeta'}_r$, due to bulk modes in the tilt-mediated $\chi$-term (Eq.~(\ref{delr_n1})). Importantly, with increasing tilt strength, rapid decay profiles resemble delta-function-like behavior around the impurity core, irrespective of the radial direction. The bulk modes contribute significantly that can lead to destructive interference away from the impurity core for the overtilted case.

\begin{figure}[h!]
	\begin{center}
		\includegraphics[width=1.0\columnwidth]{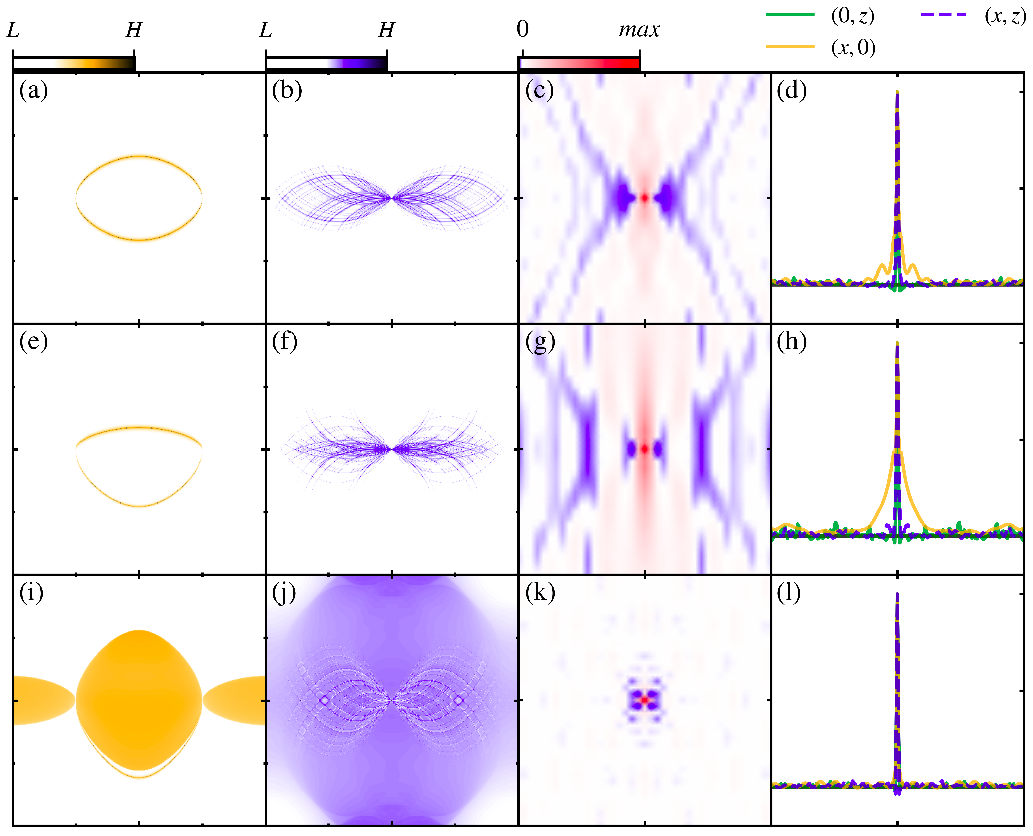}	
	\end{center}
	\vspace{-0.4cm}
	\caption{Double WSM: We repeat Fig.~\ref{fig:sWSM} for a double WSM, with the Fermi arcs are shown in (a), (e), and (i). We find a prominent leaf-like QPI profile in (b), (f),  but in (j) only around $q_x \approx 0$. The LDOS shows high anisotropy on the top surface demonstrated in (c), (g), and (k). The LDOS decays radially without showing significant oscillations, as depicted in (d), (h), and (l). We consider the same set of parameters for double WSM as considered for single WSMs in Fig.~\ref{fig:sWSM}.
	}
\label{fig:dWSM}
\end{figure} 

\subsection{Double WSMs}
\label{res2}

For double WSMs we consider $(n_0,n_x,n_y,n_z)=(t_0 \cos k_z,\cos k_x-\cos k_y, \sin k_x\sin k_y, m_z-\cos k_x-\cos k_y-\cos k_z)$, where the WNs of topological charge $n=2$ with chirality $\mp 1$ appear at ${\bm k}_c=(0,0,\pm \pi/2)$ for $m_z=2$~\cite{Fang12,yang2014classification,Roy17,Nag20}. For the untilted case, two symmetrically dispersive Fermi arcs ${\bm k}^{1,2}_{\rm FA}$ appear  (see Fig. Figs.~\ref{fig:dWSM}(a)), which lead to a leaf-like structure in the QPI (see Fig.~\ref{fig:dWSM}(b)). This reminds us strongly of a tilted single WSM. Importantly, this means that intra-Fermi arc scattering is responsible for this QPI profile, with its breadths and widths are determined by the appropriate combinations of ${\bm k}_{\rm in}\in {\bm k}^{i}_{{\rm FA}}$ and ${\bm k}_{\rm out} \in {\bm k}^{i}_{{\rm FA}}$ for a given $i=1,2$. This directly indicates the suppression of inter-Fermi arc scattering as momenta from two different Fermi arcs ${\bm k}^1_{\rm FA}$ and ${\bm k}^2_{\rm FA}$
do not mix, leading to a regular leaf-like structure only around $q_x \approx 0$-line. The suppression of inter-Fermi arc scattering is additionally evident from the tilted case. Here the asymmetric dispersive nature of Fermi arcs results in one Fermi arc ${\bm k}^1_{\rm FA}$ to be relatively flattened with respect to the other ${\bm k}^2_{\rm FA}$ (see Fig.~\ref{fig:dWSM}(e)). This causes the QPI profile around the central region close to the $q_x=0$-line to be substantially pronounced as compared to the boundary of the leaf-like region (see Fig.~\ref{fig:dWSM}(f)). As a result, the leaf-like structure is less prominent. For the overtilted type-II case, the dispersive Fermi arcs and bulk modes result in a poorly visible leaf-like structure in the center with a strong background signa lin the QPI (see Fig.~\ref{fig:dWSM}(j)).

The LDOS for the untilted case exhibits a horizontal leaf-like structure along $x$-direction around the $z=0$-line that is consistent with the QPI profile along $q_z$ in the reciprocal space (see Fig.~\ref{fig:dWSM}(c)). We notice  strong depletion of LDOS  around the diagonals $x\approx \pm z$. 
Compared to the untilted single WSM, we find much stronger suppression of LDOS away from the center along the $x$-direction (see Fig.~\ref{fig:dWSM}(d)). This we attribute this to the $1/x^l$ decay  with $l>1$ as found analytically. 
Interestingly, the rapid decay in the orthogonal direction, unlike the single WSM, is likely a consequence of 
the absence of a straight Fermi arc in double WSM. In particular, the analytical findings suggest that $\delta \rho \propto e^{i (\boldsymbol{k}_{\zeta'}- \boldsymbol{k}_\zeta)\cdot \boldsymbol{r}}/r^4$ for $\chi=0$ and $\theta^{\zeta}_r=\theta^{\zeta'}_r$ (Eq.~(\ref{delr_n2})). The above gives us a hint about the decaying behavior with the oscillatory nature of $\delta \rho$ along the $x$-direction (see Fig.~\ref{fig:dWSM}(d)).
For tilted cases, the LDOS displays a non-negligible angular profile for a short distance (see Fig.~\ref{fig:dWSM}(g)). The LDOS is significantly enhanced
near $z=0$ along $x$-direction as compared to the untilted case. This might be due to additional correction generated by the $ t_0 \exp(i\theta^{\zeta}_r)/r$ factor. 
For overtilted cases, LDOS in double WSM behaves similarly to that of a single WSM (see Fig.~\ref{fig:dWSM}(k)). 
Although the LDOS decays slower along $x$-direction for intermediate tilt as compared to the overtilted case (see Fig.~\ref{fig:dWSM}(h), (l)). 

\begin{figure}[h!]
	\begin{center}
		\includegraphics[width=1.0\linewidth]{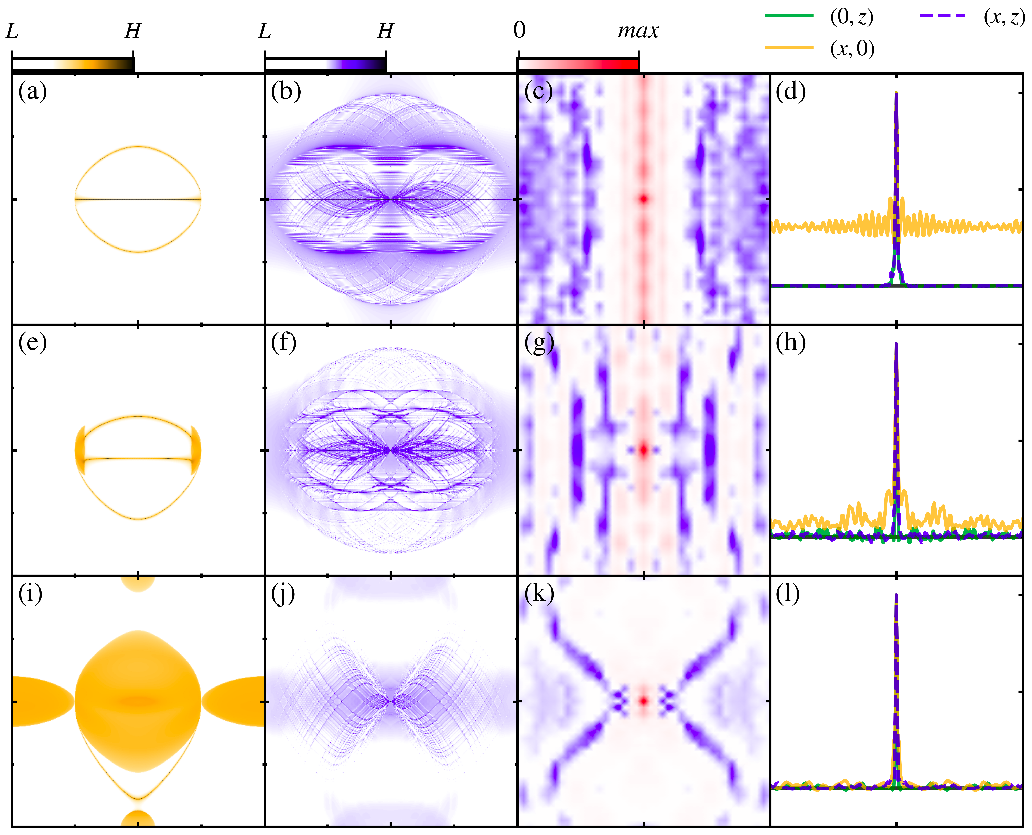}	
	\end{center}
	\vspace{-0.4cm}
	\caption{Triple WSM: We repeat Fig.~\ref{fig:sWSM} for triple WSM, where the Fermi arcs are shown in (a), (e), and (i). We find prominent oval-shaped QPI profiles in (b), (f), and (j) only around $q_x \approx 0$ that is an admixtures of straight line- and leaf-like profiles. The anisotropy in LDOS is clearly visible in (c), (g), and (k). The long-range oscillatory nature of the LDOS are depicted in (d), (h), and (l). The results are obtained with the same set of parameters as that for single WSMs in Fig.~\ref{fig:sWSM}.
 }
\label{fig:tWSM}
\end{figure} 

\subsection{Triple WSMs}
\label{res3}


For triple WSMs we consider $(n_0,n_x,n_y,n_z)=\big( t_0 \cos k_z,\sin k_x[3\cos k_y-\cos k_x -2], \sin k_y[3\cos k_x-\cos k_y -2], m_z-\cos k_x-\cos k_y-\cos k_z\big)$, where the WNs of topological charge $n=3$ with chirality $\pm 1$ appear at ${\bm k}_c=(0,0,\pm \pi/2)$ for $m_z=2$ \cite{Fang12,yang2014classification,Roy17,Nag20}. We find two dispersive Fermi arcs ${\bm k}^1_{\rm FA}$, ${\bm k}^2_{\rm FA}$, and one flat Fermi arc ${\bm k}^3_{\rm FA}$ (see Fig.~\ref{fig:tWSM}(a)). The intra-Fermi arc scattering leads to a straight and leaf-like QPI profile out of flat and dispersive Fermi arcs, respectively (see Fig.~\ref{fig:tWSM}(b)). This behavior is expected from that of single and double WSMs in the two previous subsections. Interestingly, the extended leaf-like structure along with its circular background is then an outcome of inter-Fermi arc scattering. The QPI profile for triple WSMs can be regarded to form an oval-shaped 
pattern around $q_x\approx 0$, unlike the prior leaf-like patterns. The radius of the outer circle (inner oval) shape is directly related to the scattering between the two dispersive Fermi arc ${\bm k}^1_{\rm FA}$, and ${\bm k}^2_{\rm FA}$ (straight Fermi arc ${\bm k}^3_{\rm FA}$ and dispersive Fermi arcs ${\bm k}^{1,2}_{\rm FA}$) at $k_z=0$ (see Fig.~\ref{fig:tWSM}(b)). For tilted type-I case, ${\bm k}^{1}_{\rm FA}$ (${\bm k}^{2}_{\rm FA}$) gets flattened (curved) leading to a more intense central line close to $q_x=0$ (an enlarged leaf-like region) in the QPI profile (see Figs.~\ref{fig:tWSM}(e) and (f)). The inter-Fermi arc scattering further complicates the QPI profile outside the oval-shaped region. For the overtilted type-II case, bulk modes contribute to the 
scattering in addition to the inter- and intra-Fermi arc scattering, giving rise to a complex profile around $q_x\approx 0$ as well as at $\pm \pi$ (see Fig.~\ref{fig:tWSM}(i) and (j)). However, a leaf-like QPI pattern is still visible around $q_x\approx 0$.

For the untilted case, the LDOS oscillates with an overall decaying profile along the $z$-direction around the $x = 0$ region (see Fig.~\ref{fig:tWSM}(c)). One can find an accumulation of LDOS along the $x$-direction at $z=0$ similar to the single WSM, however, there now exists spatial variation in LDOS along this direction, unlike for the single WSM. To be precise, the LDOS oscillates around a mean finite value along $x$-direction from the impurity core (see Fig.~\ref{fig:tWSM}(d)). The LDOS with such a constant average is due to the intra-Fermi arc scattering from an idealized straight Fermi arc as also noticed for single WSM in Fig.~\ref{fig:sWSM}(c). The oscillations 
 can be caused by the inter-Fermi arc scattering that is necessarily absent for a single WSM. 
Importantly, the inter-Fermi arc scattering in triple WSM results in such a notably different LDOS compared to double WSM, where LDOS exhibits algebraic decay.
The LDOS decays more rapidly along the $z$-direction as compared to $x$-direction referring to the fact that the angular LDOS profile is substantially anisotropic (see Fig.~\ref{fig:tWSM}(d)), stemming from the anisotropic Fermi arcs. The above numerical findings can be naively anticipated from the analytical findings (Eq.~(\ref{delr_n3})) such as 
$\delta \rho \propto e^{i (\boldsymbol{k}_{\zeta'}- \boldsymbol{k}_\zeta)\cdot \boldsymbol{r}})(1/r-1/r^3)$ for $\chi=0$ and $\theta^{\zeta}_r=\theta^{\zeta'}_r$. 

With increasing tilt, the LDOS diminishes substantially 
along the $x$-direction away from $z=0$-line (see Fig.~\ref{fig:tWSM}(g)). 
A careful observation suggests both the short-range and long-range decays for LDOS could be related to the predominant $1/r$ and the subdominant $1/r^3$ decay as predicted in Eq.~(\ref{delr_n3}) (see Fig.~\ref{fig:tWSM}(h)). The inter-Fermi arc scattering might cause such a profile. A long-range oscillatory profile with $r>10$ is observed for triple WSM as opposed to single and double WSMs, which we attribute to the flat Fermi arc. For the overtilted case, we observe a more extended LDOS profile, especially we find a depletion of LDOS around the diagonals $x\approx \pm z$ for triple WSM is not observed in overtilted single and double WSMs (see Figs.~\ref{fig:tWSM}(k)). Beyond this depletion the LDOS profile is much more isotropic for overtilted than less titled WSMs, reflected in the nearly isotropic LDOS profile and rapid radial decay as observed for single and double WSMs (see Figs.~\ref{fig:tWSM}(l)).

\begin{figure}
	\begin{center}
		\includegraphics[width=1.0\linewidth]{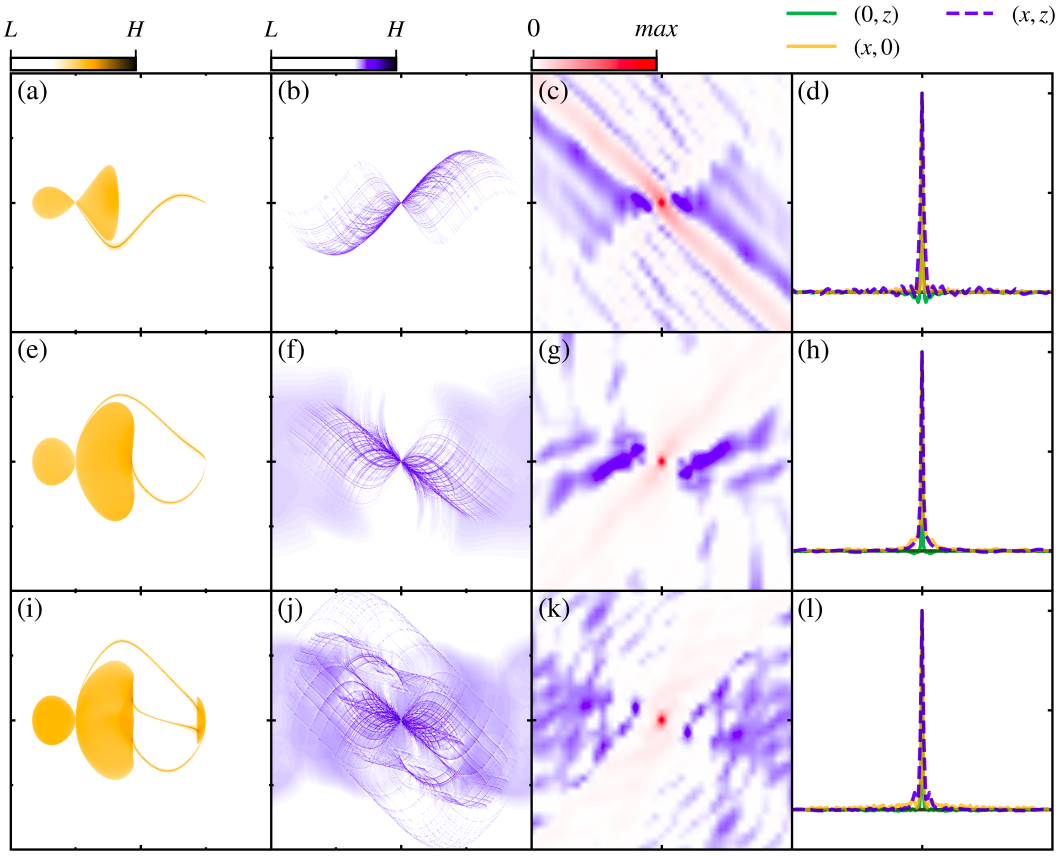}	
	\end{center}
	\vspace{-0.4cm}
	\caption{Hybrid WSMs: We repeat Fig.~\ref{fig:sWSM} for hybrid WSMs, where the Fermi arcs are shown in (a), (e), and (i). Top row show hybrid single WSM, middle row hybrid double WSM, and bottom row hybrid triple WSM.
	The Fermi arcs are in contact (not in contact) with the bulk modes for the type-II (type-I) WNs as depicted in (a), (e), and (i).
	We find rotated QPI profiles in all of the hybrid WSMs as compared to the earlier type-I and type-II WSMs. The LDOS varies diagonally along $x=\pm z$-direction, unlike the previous cases. An oscillatory profile in addition to the algebraic decay becomes visible.
	} \label{fig:hybrid}
\end{figure}

\subsection{Hybrid WSMs}
\label{res4}


We next examine impurity scattering for a hybrid phase hosting type-I and type-II WNs simultaneously \cite{Li2016, Zhang2018, Nag22}. We consider single, double, and triple WSMs with the identity term $n_0=t_1 \cos (k_z - \psi_1) + t_2 \cos (2k_z - \psi_2)$. Here the pseudospin independent first- and second-nearest neighbor hopping along $z$-direction are denoted by $t_1\exp(-i \psi_1)$ and $t_2\exp(-i \psi_2)$, with $\psi_{1,2}$ modeling a flux modulation of the hopping. Due to the hybrid nature, the Fermi arc now extends between a type-I right-side WN at ${\bm k}_c=(0,0,\pi/2)$ and type-II left-side WN at ${\bm k}_c=(0,0,-\pi/2)$ through the bulk modes (see Fig.~\ref{fig:hybrid}(a), (e), (i)). The number of Fermi arcs is again determined by the topological charge of the WNs. Note that this hybrid phase does not support any straight Fermi arc, in contrast to the untilted type-I phase.

Let us first discuss the hybrid phase for a single WSM. The non-flat dispersive Fermi arc leads to a leaf-like QPI profile. Interestingly, the non-symmetric nature of the Fermi arc with respect to $k_x$ and $k_z$ causes a deformed leaf-like profile, with its boundary around $q_x\approx q_z$
near origin $(q_x,q_z)=(0,0)$ substantially pronounced (see Fig.~\ref{fig:hybrid}(b)). The scattering from bulk modes is also present, however, we find that intra-Fermi arc scatterings are predominant. For hybrid double WSM, the deformed leaf-like structure is oriented at an angle with respect to the $q_z$-axis (see Fig.~\ref{fig:hybrid}(f)), although the angle is different compared to the hybrid single WSM. We attribute the leaf-like structure to the absence of the intra-Fermi arc scattering, which is similar to the previous cases with single and double WSMs. 
However, we note that the leaf-like structure notably rotates in momentum space, in contrast to the previous type-I and type-II phases showing a leaf-like structure formed along the $q_z$-axis around $q_x\approx 0$. For hybrid triple WSM, in addition to the intra-Fermi arc scattering, inter-Fermi arc scattering leads to finite QPI intensity also along the $q_x$-axis at $q_z=0$ (see Fig.~\ref{fig:hybrid}(j)). A rotated oval-shaped QPI profile along the axis $q_x=-q_z$ is observed with finite scattering momenta $q_x\ne 0$ around $q_z \approx 0$. The bulk modes also participate in building up the background profile for QPI, more and more prominently with increasing topological charge.

We remarkably find an accumulation of LDOS around the diagonal along $x=-z$-($x=z$-)direction for hybrid WSMs with topological charge unity (two and three) (see Fig.~\ref{fig:hybrid}(c,g,k)). And we find a strong suppression of LDOS along the $x$-direction around $z=0$-line for all three instances. 
This is caused by the QPI profiles in momentum space that are more pronounced along the diagonal orthogonal to the real space LDOS profile. A similar orthogonality was seen for the non-hybrid WSMs. However, the LDOS structure obtained from the previous non-hybrid WSMs instead showed depletion along the diagonals. This difference is due to the fact that one edge of the leaf-like structure is more pronounced than the other edge for hybrid WSMs, while they are equally pronounced for the non-hybrid WSMs. We also note that the resulting orientation of the QPI profiles is opposite between single and double WSMs in the hybrid phase, which can be traced back to the different curvatures of the Fermi arcs.
With increasing the topological charge further, the amplitude of the diagonal LDOS becomes suppressed (see Fig.~\ref{fig:hybrid}(k)). Some oscillations in the LDOS exists, but it  quickly decays with radial distance, corroborating theoretical prediction for tilted WSMs, in hybrid WSMs irrespective of the topological charge (see Fig.~\ref{fig:hybrid}(d), (h), (l)).

\begin{figure*}[ht]
	\begin{center}
		\includegraphics[width=0.48\linewidth]{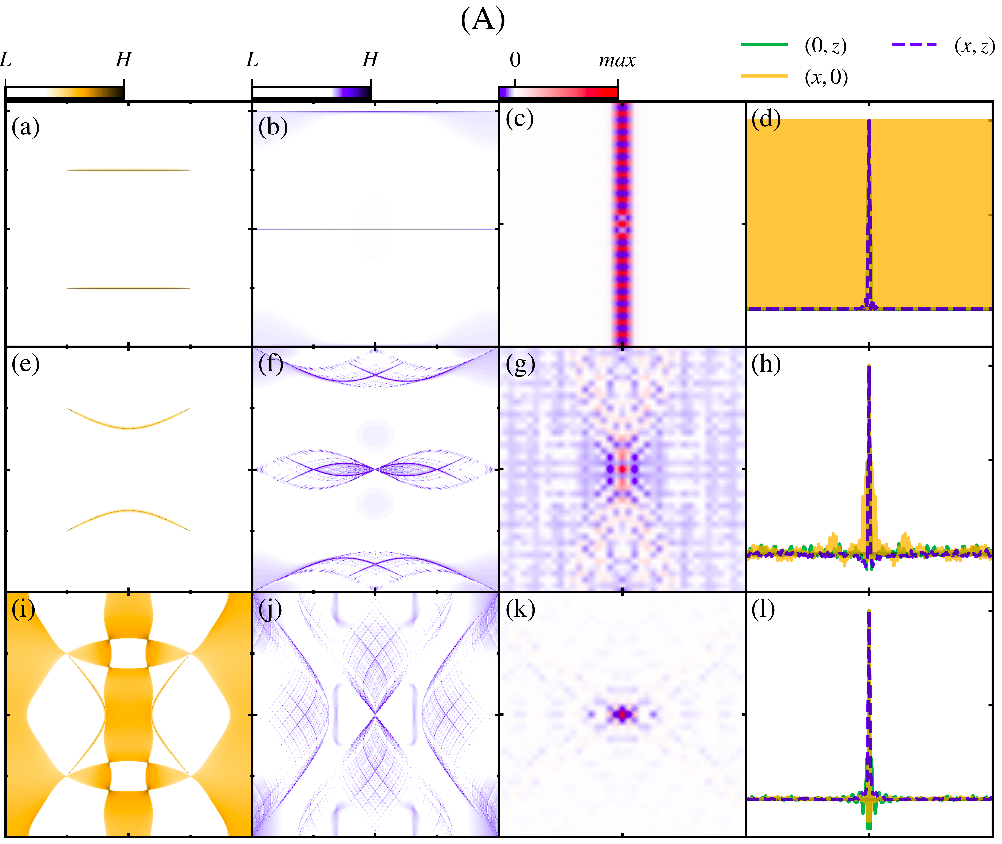}
 \includegraphics[width=0.48\linewidth]{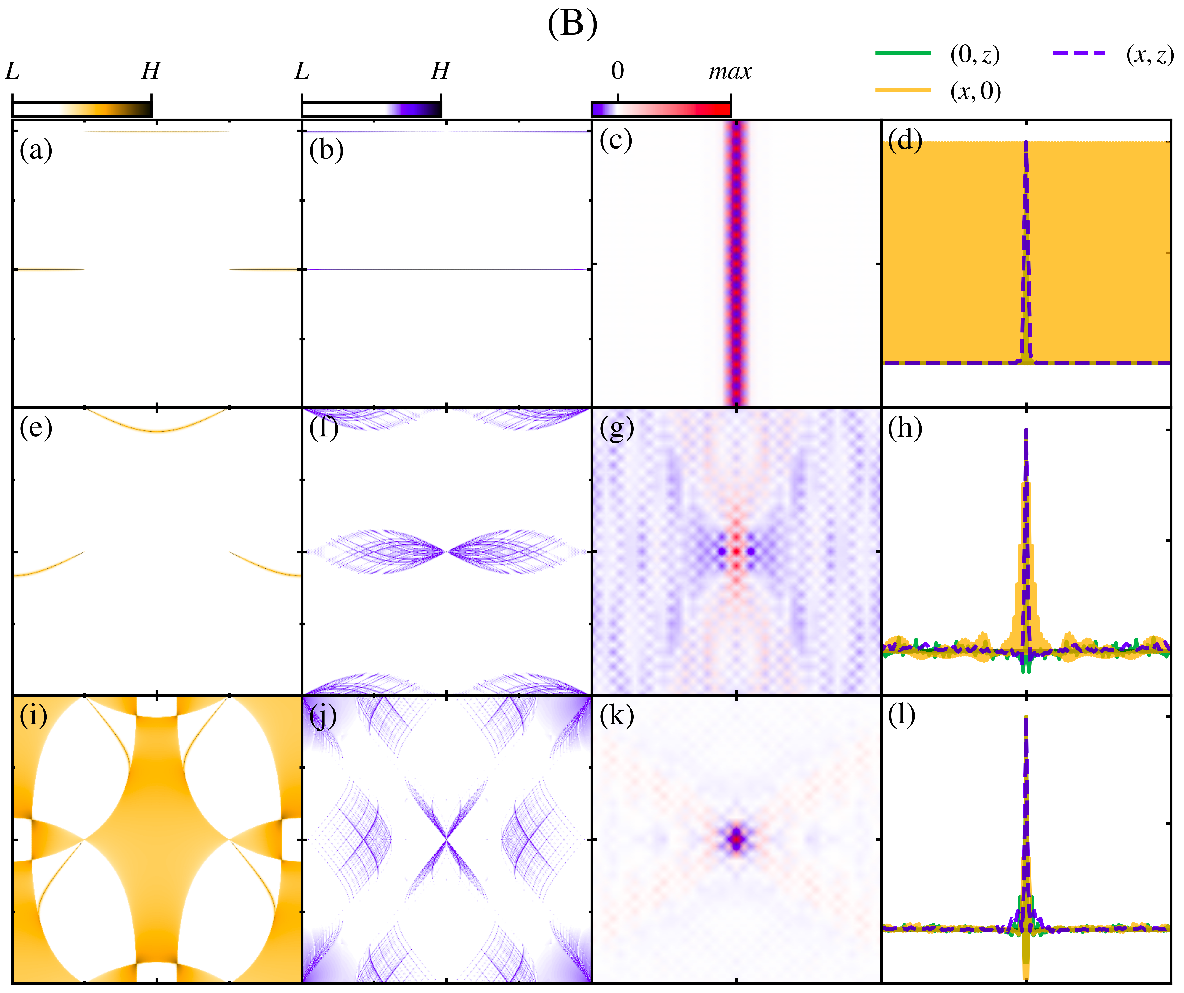} 
	\end{center}
	\vspace{-0.4cm}
	\caption{TRS invariant single WSM (A) and TRS broken single WSM (B): We repeat Fig.~\ref{fig:sWSM} for for TRS invariant and broken single WSMs, where the Fermi arcs are shown in (a), (e), and (i). 
	Top row shows untilted case with $t_0=0$, middle row moderate tilted case with $t_0=0.5$, and bottom row overtilted case with $t_0=1.2$.
	We find one Fermi arc out of two lying within (across) the BZ for TRS invariant (broken) WSM. The QPI profile around $q_x\approx \pm \pi$ for the tilted cases in (A-f,j) can be distinguished from (B-f,j). The angular nature of LDOS and radial decay profiles are also different as shown in (A-g,k), (B-g,k) and (A-h,l), (B-h,l). We find rapid radial oscillations in LDOS around a finite (vanishingly small) averaged value referring to the presence of inter-Fermi arc scattering for untilted (tilted) cases. 
	} \label{fig:sWSMTIbroken}
\end{figure*}

\subsection{TRS invariant and broken pure single WSMs}
\label{res5}


We next compare TRS invariant and broken WSMs. For the former we consider $(n_0,n_x,n_y,n_z)=(t_0 \cos k_z,1-\cos^2 k_x-\cos k_y-\cos k_z, \sin k_y, \cos k_z)$ that respects TRS, generated by ${\mathcal K}$ (complex conjugation) but breaks IS, generated by $\sigma_x$, leading to two pairs of WNs at ${\bm k}_c=(\pm \frac{\pi}{2},0,\pm \frac{\pi}{2})$ 
\cite{McCormick17}. For the TRS broken case, in order to match with the four WNs with the TRS invariant case, we consider the TRS broken single WSM, already described previously, but now with $m_z=0$. This results in four WNs at ${\bm k}_c=(0,\pi,\pm\frac{\pi}{2})$ and ${\bm k}_c=(\pi,0,\pm\frac{\pi}{2})$.  Note that there were two WNs for $m_z=2$  used in Fig. \ref{fig:sWSM}. 

Having established the same number of WNs for both the TRS invariant and broken case, we can compare their scattering responses. First we plot their Fermi arcs in the untilted case (see Figs.~\ref{fig:sWSMTIbroken}(A-a) and (B-a)). We find two Fermi arcs of opposite chiralities for both models.
For the TRS-broken WSM, we find that one of the Fermi arcs is connected across the BZ, unlike the TRS invariant case where both the Fermi arcs are connected within the BZ. To be precise, we find ${\bm k}^{1[2]}_{{\rm FA}}=(k_x=\pi/2[-\pi/2],-\pi/2<k_z<\pi/2)$ for the TRS invariant case and 
${\bm k}^1_{{\rm FA}}=(k_x=0,\pi/2<k_z<3\pi/2)$, ${\bm k}^2_{{\rm FA}}=(k_x=\pi,-\pi/2<k_z<\pi/2)$ for the TRS broken case.
The QPI profiles for both these models look qualitatively similar around $q_x \approx 0$, as expected when ${\bm k}_{\rm in} $ and ${\bm k}_{\rm out}$ share the same set of values of $k_x$. On the other hand, the QPI profiles look different around $q_x \approx \pi$ as ${\bm k}_{\rm in} $ and ${\bm k}_{\rm out}$ contains different set of $k_x$ separated by $\pi$. Based on the results for singel WSMs, we attribute the above characters to intra- and inter-Fermi arc scattering for $q_x \approx 0$ and $q_x \approx \pi$, respectively (see Fig.~\ref{fig:sWSMTIbroken}(A-b) and (B-b)). 
The difference at $q_x \approx \pi$ for the two cases is that the QPI intensity vanishes at $(q_x,q_z)=(\pi,0)$ only for TRS broken case, due to an absence of scattering between the surface projection at ${\bm k}_c=(0,\pm \pi/2)$ and ${\bm k}_c=(\pi,\pm \pi/2)$ of bulk WNs with the same chirality. This prohibition of scattering between the same chirality WNs also causes the vanishing profile of the QPI at ${\bm q}=(0, \pm \pi)$ [${\bm q}=(\pi, \pm \pi)$] for both the models [only TRS invariant model]. 

With increasing tilt, the Fermi arcs bend identically (oppositely) for TRS broken (invariant) WSMs [see Fig.~\ref{fig:sWSMTIbroken}(A-e) and (B-e)]. This causes the recurrence (disappearance) of the leaf-like pattern around $q_x\approx \pm \pi$ for TRS broken (invariant) case (see Fig.~\ref{fig:sWSMTIbroken}(A-f) and (B-f)). 
To be precise, the QPI intensity appears (vanishes) over the line $q_x=\pm \pi$ for TRS broken (invariant) WSM. Instead the TRS invariant case shows intensity at smaller but finite $q_x$.
The vanishing intensity of the QPI profile at $(0,\pm \pi)$ and $(\pm \pi,0)$ is similarly observed for the tilted case as well.
For the overtilted case where Fermi arcs terminate into the bulk, the QPI instead exhibits a vertical leaf-like structure for the TRS invariant case more prominently as compared to TRS broken case (see Fig.~\ref{fig:sWSMTIbroken}(A-j) and (B-j)).
The leaf-like QPI profile is not always observed 
for curved Fermi arc doublet when TRS is preserved (see Fig.~\ref{fig:sWSMTIbroken} (A-f)); this is in contrast to  Fig. \ref{fig:sWSM} where
the leaf-like QPI pattern appear in presence  of a
single curved Fermi arc and in absence of TRS.

Now turning to the spatial LDOS profile, we find that TRS invariant and broken WSMs behave in a qualitatively similar manner for all tilts (see Fig.~\ref{fig:sWSMTIbroken}(A-c,g,k) and (B-c,g,k)).  However, there exist some  differences. 
This is in contrast to the momentum-space QPI profiles that behave quite distinctly for the above two models. 
With increasing the tilt, the LDOS profile builds up away from the $z=0$-line for both models referring to the fact that LDOS behavior is not sensitive to the 
bending characteristics of Fermi arcs. 
For the untilted case, we find rapid oscillations in $x$ over the 
$z=0$-line which we attribute to  the inter-Fermi arc scattering between the 
straight Fermi arcs (see Fig.~\ref{fig:sWSMTIbroken}(A-d) and (B-d)). Note that the LDOS oscillates around a 
 constant finite average that could be considered as a signature of intra-Fermi arc scattering from the idealized straight Fermi arc also found for single WSM in Fig.~\ref{fig:sWSM}.  The oscillation period is related to $q_x \approx \pi$ while the mean average value might be related to $q_x \approx 0$. 
Interestingly, when the tilt strength is moderate, the LDOS decays rapidly close to the impurity core while a slower decay is observed for a larger distance (see Fig.~\ref{fig:sWSMTIbroken}(A-h) and (B-h)). 
This double decay nature may be related to the inter-Fermi arc scattering where $1/r^3$ and $1/r$ decay are obtained analytically in Eq.~(\ref{delr_n3}) for the radial decay profile in a triple WSM. However, the clear evidence of $1/r^3$ decay is yet to be examined in the future, as is the applicability to TRS invariant WSMs.
Long-distance oscillations are observed here similar to the triple WSMs as shown in Fig.~\ref{fig:tWSM}(d) indicating the 
presence of the inter-Fermi arc scattering between the curved Fermi arcs. 
For the overtilted type-II case, the LDOS decay quite rapidly 
similar to the type-I phase, however, the oscillations are significantly suppressed as compared to other tilted type-I cases (see Fig.~\ref{fig:sWSMTIbroken}(A-l) and (B-l)). Comparing to Fig.~\ref{fig:sWSM} with a single Fermi arc, we find  repeated leaf-like patterns in Fig.~\ref{fig:sWSMTIbroken} (B) due to the Fermi arc doublet while both the models break TRS.

\begin{figure}
	\begin{center}
		\includegraphics[width=1.0\linewidth]{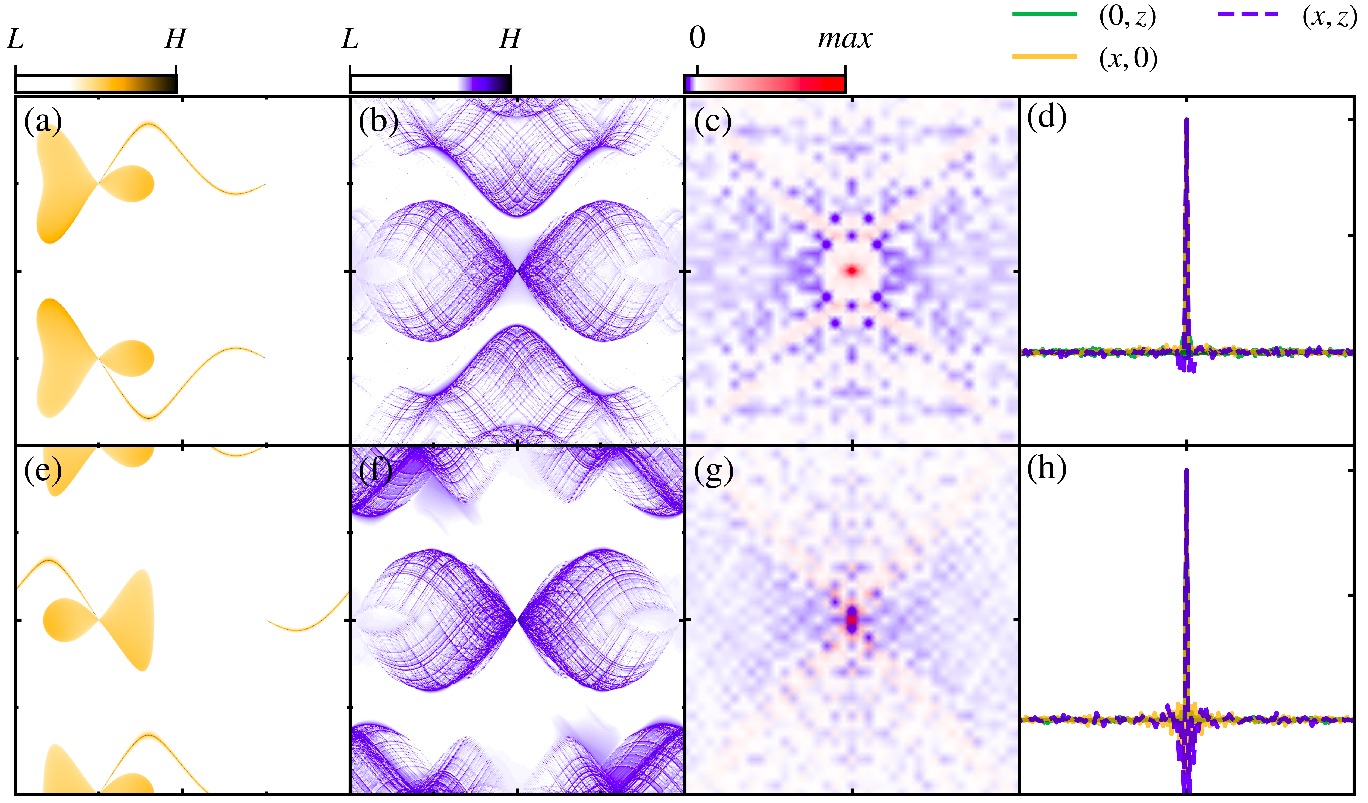}	
	\end{center}
	\vspace{-0.4cm}
	\caption{TRS invariant and broken hybrid single WSM: We repeat Fig.~\ref{fig:sWSM} for TRS invariant (top row) and broken hybrid single WSMs (bottom row).
	The Fermi arc connectivities are different for these two models as shown in (a) and (e). The QPI profiles around $q_x\approx \pm \pi$, the LDOS structure around the impurity core, and the radial decay characteristics also vary for these two cases. 
	} 
	\label{fig:sWSMTIbroken_hy}
\end{figure}

\subsection{TRS broken and invariant hybrid single WSMs}
\label{res6}


Finally, we examine the hybrid case when considering the TRS broken and invariant WSMs discussed in the last subsection. For this purpose, we replace $t_0 \cos k_z \sigma_0\rightarrow [t_1 \cos (k_z - \psi_1) + t_2 \cos (2k_z - \psi_2)]$, with $t_1=t_2=0.5$ and $\psi_{1,2}=\pi,\pi/2$ in the previously discussed hybrid single WSM in Section~\ref{res4}.
For the TRS invariant WSM, the Fermi arcs, connecting WNs at ${\bm k}_c=(\pi/2,0, \pm \pi/2)$ and ${\bm k}_c=(-\pi/2, 0,\pm \pi/2)$, bend oppositely with respect to each other (see Fig.~\ref{fig:sWSMTIbroken_hy}(a)). The bulk modes are also visible around $k_x =\pm \pi/2$ with $k_z=-\pi/2$.
The Fermi arcs for TRS broken WSM, connecting WNs at ${\bm k}_c=(0, \pi, \pm \pi/2)$ and ${\bm k}_c=(\pi,0, \pm \pi/2)$, bends identically (see Fig.~\ref{fig:sWSMTIbroken_hy}(e)). The bulk modes are also visible around $k_x =0,\pi$ with $k_z=-\pi/2$. The QPI profiles in the hybrid phase for both cases closely resemble their behavior in the type-I phase with moderate tilt as shown in Fig.~\ref{fig:sWSMTIbroken}(A-f) and (B-f).

A leaf-like structure appears around $q_x \approx 0$ for both the models, while around $q_x\approx \pm \pi$, the QPI shows different patterns for 
TRS invariant and broken hybrid WSMs (see Fig.~\ref{fig:sWSMTIbroken_hy}(b) and (f)). This profile is mainly caused by the scattering from the bulk modes, connected with two Fermi arcs in addition to the scattering between the Fermi arc surface modes. 
Moreover, the QPI profile vanishes around $q_z\approx \pm \pi$ ($q_z\approx 0$) for $q_x\approx \pm \pi$ in the case of TRS invariant (broken) hybrid WSM. This we attribute to the absence of the inter-Fermi arc scattering between the same chirality WNs. A close inspection suggests that QPI profile becomes  convex (concave) oscillatory around $q_x\approx \pm \pi$ unlike the central leaf-like pattern around $q_x\approx 0$ for TRS invariant (broken) hybrid single WSM.

We find a substantial variation in the LDOS along the diagonals $x=\pm z$ for both cases, but, the amplitudes of the LDOS decay quicker for TRS invariant case compared to the broken counterpart (see Fig.~\ref{fig:sWSMTIbroken_hy}(c), and (g)). Such an angular variation is dominated by the type-II nature of the WNs and the scattering between the bulk modes. When compared with the hybrid single WSM with broken TRS, shown in Fig.~\ref{fig:hybrid}(c), we find prominent LDOS structure only along $x=-z$ due to the intra-Fermi arc scattering within a single curved Fermi arc. In contrast, in the present case, inter-Fermi arc scattering between two different curved Fermi arcs leads to a double diagonal structure of LDOS. The LDOS decays very fast along the $x$- and $z$-directions for the TRS invariant case as compared to the TRS broken hybrid single WSM (see Fig.~\ref{fig:sWSMTIbroken_hy}(d), and (h)). We additionally find different angular and radial variations of LDOS around the impurity for these two models that are evident from the microscopic LDOS distribution shown (see  Figs.~\ref{fig:sWSMTIbroken_hy}(c) and (g)). 

\section{Discussion, Conclusion, and outlook}
\label{conclusion}


In this work we consider a localized surface impurity in order to analyze the structure of the Fermi arc in $(k_x,k_z)$-plane to provide signatures to distinguish a range of different WSMs. In particular, we consider single, double, and triple WSMs, type-I and type-II, as well as hybrid WSMs, and also both TRS invariant and broken WSMs. We primarily focus on the QPI profile, obtained using the $T$-matrix formalism, in momentum space spanned by $(q_x,q_z) \in {\bm k}_{\rm in}-{\bm k}_{\rm out}$.
In other words, our study aims at extracting the Fermi arc momenta out of intriguing QPI profiles, and thereby also the type of WSM. We also examine the real-space modulation of LDOS on the top surface that is obtained by Fourier transforming the QPI profile.  To be precise, 
our study reveals the angular and radial $(\theta, r)$ variations of LDOS away from the impurity. 

As a first step, we  study single, double, and triple WSMs where the number of Fermi arcs increases with topological charge $n = 1, 2, 3$, respectively. Here we are able to partly analyze the LDOS profile analytically, considering an approximate low-energy Hamiltonian. We find a universal $1/r$ decay as a signature of the
WSM, where there exist additional $1/r^n$ corrections for $n>1$ for topological charge greater than unity (see Eqs.~(\ref{delr_n1}), (\ref{delr_n2}), and (\ref{delr_n3})). By also studying a more accurate lattice model of these WSMs, we find that inter-Fermi arc scattering appears only for triple WSM, while only intra-Fermi arc scattering is present for single and double WSMs (see Figs. \ref{fig:sWSM}, \ref{fig:dWSM}, and \ref{fig:tWSM}). We might therefore qualitatively attribute inter-Fermi arc scattering to the special
$1/r^3$-decay for LDOS in triple WSM found analytically. On the other hand, 
intra-Fermi arc scattering can be captured by $1/r$-decay. These decay profiles are in addition associated with a periodic angular dependence. 

Moving on to the momentum space profile, which we extract numerically, we find that the QPI acquires minimum intensity at $(q_x,q_z)=(0,2 k^{p}_z)$ indicating the surface projection of WNs at $(k_x,k_z)=(0,\pm k^{p}_z)$, regardless of the topological charge or the tilt of type-I WSM. Furthermore, a straight Fermi arc leads to the straight-line feature in QPI, while  curved ones give rise to a leaf-like feature, with its vertical width and horizontal length reflecting the maximum momenta dispersed for the Fermi arcs along $k_x$- and $k_z$-direction, respectively. Importantly, the inter-Fermi arc scattering leaves its marked signature along $q_x$ around $q_z \approx 0$ that is clearly present (absent) for triple (single and double) WSM(s). The vertical QPI structure along $q_x$ away from $q_z=0$ can also considered as an identification for type-II WSMs as compared to type-I WSMs. 
Finally, for overtilted type-II cases, the bulk modes add a strong background signal on top of the above QPI structure from the Fermi arcs. 
Interestingly, for the numerically extracted impurity-induced change in the LDOS, we observe a non-zero LDOS along the $x$-direction, clearly noticed for the untilted single WSM, where only one straight Fermi arc exists along $k_z$. This LDOS starts to oscillates along $x$-direction around a finite mean value, as noticed for untilted triple WSM, where straight as well as curved Fermi arcs both are present. A characteristically different variation in the LDOS along $x$-direction is found for untilted double WSM where there exist no straight Fermi arcs. We can hence anticipate that a straight Fermi arc causes a constant (oscillatory) LDOS in the transverse direction in the absence (presence) of inter-Fermi arc scattering. The above special behavior can be naively 
understood from the spatial dependence of the LDOS, considering the phase factor $e^{i (\boldsymbol{k}_{\zeta'}- \boldsymbol{k}_\zeta)\cdot \boldsymbol{r}}$ with $\theta^{\zeta}_r=\theta^{\zeta'}_r$, $\chi=0$ and ignoring the radial decay, that are obtained analytically in Eqs.~(\ref{delr_n1}), (\ref{delr_n2}), and (\ref{delr_n3}). 
However, we note that further studies are required in order to understand the accurate decay nature of LDOS and connect them with the intra- and inter-Fermi arc scattering.

\begin{table*}[ht]
\begin{center}
\begin{tabular}{| p{1.9cm}|p{1.7cm}|p{1.7cm}|p{1.7cm}|p{4.4cm}|p{4.4cm} |} 
\hline
 WSM & Number of WNs and charge  & Number of Fermi arcs & Inter-Fermi arc scattering  & QPI profile in  presence of tilt (apart from background blur)&  Radial LDOS profile in presence of tilt (apart from algebraic decay) \\ 
\hline \hline
 Single WSM (Fig.~\ref{fig:sWSM})   & $2$, $\pm 1$ &  $1$ & no & type-I: uniform leaf-like, type-II: X-shape near  origin  & type-I: short-range oscillations, type-II: no short-range oscillations   \\ 
\hline
 Double WSM (Fig.~\ref{fig:dWSM})    & $2$, $\pm 2$ &  $2$ & no & type-I: nonuniform leaf-like, type-II: uniform leaf-like  & type-I and type-II: no short-range oscillations   \\ 
\hline
Triple  WSM (Fig.~\ref{fig:tWSM})   & $2$, $\pm 3$ &  $3$ & yes & type-I: inner over-shaped outer circle-like, inter Fermi arc scattering, type-II: weak leaf-like   & type-I: oscillations around finite mean, type-II: substantially suppressed oscillations around zero mean     \\ 
\hline 
 Hybrid WSM ($n=1,2,3$) (Fig.~\ref{fig:hybrid})   & $n$, $\pm n$ &  $n$ & only $n=3$ & $n=1$: nonuniform  leaf-like, $n=2$: nonuniform tilted leaf-like, $n=3$: tilted over-shaped   & $n=1,2$: negligible short-range oscillations, $n=3$: substantially suppressed short-range oscillations   \\ 
 \hline 
 TRS invariant single WSM (Fig.~\ref{fig:sWSMTIbroken}(A))   & $4$, $\pm 1$ &  $2$ & yes & type-I: nonrepeated leaf-like, type-II: X-shape near origin and presence of vertical leaf-like  & type-I: rapid short-range oscillations, type-II: substantially suppressed short-range oscillations   \\ 
  \hline 
 TRS broken single WSM (Fig.~\ref{fig:sWSMTIbroken}(B))   & $4$, $\pm 1$ &  $2$ & yes & type-I: repeated leaf-like, type-II: X-shape near origin and absence of vertical leaf-like  & type-I: rapid short-range oscillations, type-II: substantially suppressed short-range oscillations   \\ 
\hline 
 TRS invariant hybrid single WSM (Fig.~\ref{fig:sWSMTIbroken_hy}, top row)   & $4$, $\pm 1$ &  $2$ & yes & central leaf-like and boundary convex-oscillatory  & substantially suppressed short-range oscillations   \\ 
 \hline 
 TRS broken hybrid single WSM (Fig.~\ref{fig:sWSMTIbroken_hy}, bottom row)   & $4$, $\pm 1$ &  $2$ & yes & central leaf-like and boundary concave-oscillatory   & suppressed short-range oscillations   \\ 
\hline
\end{tabular}
\end{center}
\caption{Summary of main findings on QPI and LDOS profiles in different WSMs.}
\label{tab:Bz} 
\end{table*}

As a next step we extend our analysis to the hybrid phase where one WN is type-I and its chiral partner is type-II such that pocket-like and point-like 
Fermi surfaces coexist (see Fig. \ref{fig:hybrid}). In this case we find a leaf-like structure due to the curved Fermi arc. However, a uniform leaf-like QPI profile is not observed as for the non-hybrid cases, rather it is substantially strong only along one edge around $q_x\approx q_z$. This is markedly different  compared to the non-hybrid WSMs, where the leaf-like structure is uniformly visible. 
With increasing topological charge, Fermi arcs of opposite curvature contribute to the rotated leaf-like QPI profile. Therefore, hybrid double and triple WSM can be distinguished from their non-hybrid counterparts as far as the orientation of the leaf-like structure is concerned. Again, we find signs of inter-Fermi arc scattering only for hybrid triple WSM, as evident from the QPI profile along $q_x$ around $q_z \approx 0$. 
Interestingly, the corresponding LDOS builds up diagonally in the hybrid phases, instead of vertically for the non-hybrid cases. However, higher topological charge leads to a rapid decay as consistent with the analytical findings. 

We then turn to QPI profiles for TRS invariant and broken WSMs. Here we compare cases where they both have four WNs (see Fig. \ref{fig:sWSMTIbroken}). The absence of scattering between the same chirality WNs, residing on two different Fermi arcs, leads to a unique QPI profile in the $q_x=\pi$ region. This allows us to track the position and chirality of the WNs in the BZ. A tilt causes the leaf-like profile in the QPI to show up differently around $q_x=\pm \pi$ and $q_z=\pm \pi$ for the TRS invariant and broken WSMs. Interestingly, for the TRS broken single WSM with two WNs, the leaf-like QPI profile only appears around $q_x \approx 0$, thus rendering a qualitative difference for TRS broken WSMs with different number of WNs. The LDOS for the untilted WSMs behaves similarly, irrespective of their TRS broken and invariant nature. This can be understood qualitatively from the low-energy surface Hamiltonian derived analytically around an isolated WN. 
However, we can distinguish TRS invariant from broken WSMs with tilt by the microscopic details of the LDOS structure. Additionally, we analyze the TRS broken and TRS invariant hybrid WSMs where two WNs are of type-II and the other two correspond to type-I (see Fig. \ref{fig:sWSMTIbroken_hy}). The QPI profiles around $q_x=\pm \pi$ are distinct, caused by the inter-Fermi arc scattering, for the two types of WSMs. The leaf-like profile around $q_x= 0$ is the signature of scattering within the same curved Fermi arc. The diagonal variation of LDOS provides another distinct signature. Both WSMs show inter-Fermi arc scattering that can be responsible for rapid radial decay, however, this needs to be analyzed in the future with great detail. 

Having demonstrated the intra- and inter-Fermi arc scattering in many different models, we commonly find that the 
latter can be present for IS broken as well as in preserved cases. For example, pure and hybrid double WSM models (see Figs.~\ref{fig:dWSM}, and \ref{fig:hybrid}), considered here, break the IS i.e., $Ph(\boldsymbol{k})P^{-1}\ne h(-\boldsymbol{k})$ with $P=\sigma_z$ and do not support inter-Fermi arc scattering. By contrast, pure and hybrid triple WSMs both exhibit inter-Fermi arc scattering while the former (latter) respects (breaks) IS. Notice that all the hybrid cases we consider break IS due to the identity term, while pure single and triple WSMs preserve IS. Therefore, we find that inter-Fermi arc scattering is not directly related to the IS of the model, rather the details of the band structure might be responsible for such scattering. It would be an interesting future direction to more in detail investigate the origin of inter-Fermi arc scattering, for example by considering other models. 

We also need to point out that in our work we are occasionally considering Fermi arcs that  are flat at zero-energy and non-dispersive, which is clearly an idealized situation. Problematic is that we find that such straight Fermi arcs cause a constant impurity-induced LDOS profile or oscillatory LDOS  around a non-zero mean value depending on the specific location of the Fermi arc. Such long-range LDOS response is not a realistic for a localized impurity. Note that we here only investigate the impurity-induced LDOS at zero-energy. We believe that there will be a depletion of LDOS from finite energy to at least compensate in terms of total amount of states in the system.
Interestingly, finite tilt always yields non-flat dispersive Fermi arc in our models, and there we always find that the LDOS decays spatially away from the impurity. We therefore attribute the non-decaying LDOS to the pathologically flatness of the Fermi arc. Since Fermi arcs in real materials are always at least slightly dispersive, we do not expect this idealized, pathological case of non-decaying LDOS to be experimentally present. Still, our results highlight that for very flat Fermi arcs, very long-range impurity-induced LDOS changes are expected. \textcolor{black}{In this context, we also note that the analytical technique adopted here is an effective approach to obtain a closed-form expression for the radial decay of that LDOS that can qualitatively hint towards it basic features. However, more detailed microscopic analytical calculations can be implemented in the future to estimate the quantitative features.}

\textcolor{black}{Having discussed various findings on different WSMs, we now also discuss the connections and limitations of $T$-matrix approach, as far as the real experimental results are concerned. First of all, we note that  $T$-matrix approach is not a perturbative approach \cite{Balatsky2006} and we thus do not neglect any correction terms that may alter the QPI profiles. Overall, the $T$-matrix approach is valid if the impurities are sparse and they do not interact among themselves i.e., without any long-range correlation among each other. Thus for clean sample with only sparse impurities to generate the QPI profiles, our results should be experimentally relevant \cite{hanaguri2007quasiparticle,sharma2020momentum,hirschfeld2015robust}.
Moreover, our QPI results from model calculations of a single WSM closely resembles that of a recent first principle calculation \cite{Kourtis16}, further supporting the validity of our results. In particular we point out that different types of WSMs, exhibiting various orientations and natures of Fermi arcs and bulk WNs structures, can be experimentally distinguished as long as the dilute impurity limit is applicable. For more dirty samples we expect the leaf-like QPI profiles in Figs.~\ref{fig:sWSM}-\ref{fig:sWSMTIbroken_hy}  to get deformed and randomly weighted with contributions coming from the otherwise white regions due to complex scattering processes. 
Finally, our results are also more accurate when the sample size is relatively large such that in-plane momenta are well-defined. 
We have checked our main results using larger samples with a mesh up to size $800 \times 800$ and find no significant corrections. Taken together, we expect that our lattice results are robust and experimentally accurate in large sample with low defect concentrations. 
}

To summarize, our systematic investigations of momentum space QPI and real space LDOS shed light on the characteristics of surface Fermi arcs in the WSMs. We summarize our results in Table.~\ref{tab:Bz}. 
Given the fact that our results are based on tight-binding models, we believe these QPI profiles can directly be verified by the STM experiments~\cite{zheng2016atomic,Hao2018}. 
One complication is that there often exist multiple WNs in most real WSMs materials. Our findings are directly relevant when two and/or four WNs  are present close to the Fermi energy, but some conclusions can also be drawn for more WNs based on our results. Our study can also be extended to Dirac or nodal semimetal to probe the surface states, with straightforward additional calculations.
We have here considered only spin-unresolved QPI and its associated LDOS structures for WSMs, and the results are thus insensitive to the magnetic nature of the impurity potential for all WSMs that already break TRS. However, the effect of magnetic impurity on the TRS invariant WSMs is an open future question. 
Moreover, the effect of a surface impurity on the bulk physics of WNs is another interesting point to investigate in presence of external magnetic and electric fields.  In particular, the connection between the surface Fermi arc and the bulk  chiral anomaly effect in presence of  impurity may be interesting, but clearly go beyond our work.

\acknowledgements
\label{ack}
We thank Carsten Honerkamp for insightful discussion on the project. 
FX thanks the German Science Foundation (DFG) for support sincerely
through RTG 1995 and RWTH0662 for granting computing time.
TN and ABS thank the Swedish Research Council (Vetenskapsr\aa det Grant No.~2018-03488), the Knut and Alice Wallenberg Foundation through the Wallenberg Academy Fellows program and project grant KAW 2019.0068.

\bibliography{qpi_mWSM.bib}

\end{document}